\documentclass[useAMS,usenatbib,usegraphicx]{mn2e}
\usepackage{appendix}
\usepackage{graphicx}  % Include figure files
\usepackage{dcolumn}   % Align table columns on decimal point
\usepackage{subfigure}

\usepackage{color}
\usepackage{graphicx}
\usepackage{makeidx}
\usepackage{epsfig}
\usepackage{epstopdf}
\usepackage{natbib}
\usepackage{float}
\usepackage{amsmath}
\usepackage{times}
\usepackage{upgreek}
\usepackage[varg]{txfonts}
\usepackage{enumerate}
\usepackage{verbatim}
\usepackage{booktabs}
\usepackage{multirow}
\usepackage{amssymb}
\usepackage{placeins}
\usepackage{array,longtable}
\usepackage{amsfonts}
\usepackage{subfigure,amsmath,amssymb ,amsfonts, latexsym}
\usepackage[notcite,color,final]{showkeys}

\title[]
{Assessing the effect of lens mass model in cosmological application with updated galaxy-scale strong gravitational lensing sample}

\author[Chen et al.]
{Yun Chen$^{1}$\thanks{E-mail:chenyun@bao.ac.cn}, Ran Li$^{1,2}$, Yiping Shu$^{3}$, Xiaoyue Cao$^{1,2}$\\
$^1$ Key Laboratory for Computational Astrophysics, National Astronomical Observatories, Chinese Academy of Sciences, Beijing, 100101, China;\\
$^2$ University of Chinese Academy of Sciences, 19 A Yuquan Rd, Shijingshan District, Beijing, 100049, China;\\
$^3$ Institute of Astronomy, University of Cambridge, Madingley Road, Cambridge CB3 0HA, UK}

\begin{document}

\date{\today}

\voffset- .5in

\pagerange{\pageref{firstpage}--\pageref{lastpage}} \pubyear{}

\maketitle

\label{firstpage}

\begin{abstract}
By comparing the dynamical and lensing masses of early-type lens galaxies, one can constrain both the cosmological parameters and the density profiles of galaxies. We explore the constraining power on cosmological parameters and the effect of the lens mass model in this method with 161 galaxy-scale strong lensing systems, which is currently the largest sample with both high resolution imaging and stellar dynamical data. We assume a power-law mass model for the lenses, and consider three different parameterizations for $\gamma$ (i.e., the slope of the total mass density profile) to include the effect of the dependence of $\gamma$ on redshift and surface mass density. When treating $\delta$ (i.e., the slope of the luminosity density profile) as a universal parameter for all lens galaxies, we find the limits on the cosmological parameter $\Omega_m$ are quite weak and biased, and also heavily dependent on the lens mass model in the scenarios of parameterizing $\gamma$ with three different forms. When treating $\delta$ as an observable for each lens, the unbiased estimate of $\Omega_m$ can be obtained only in the scenario of including the dependence of $\gamma$ on both the redshift and the surface mass density, that is $\Omega_m = 0.381^{+0.185}_{-0.154}$ at 68\% confidence level in the framework of a flat $\Lambda$CDM model. We conclude that the significant dependencies of $\gamma$ on both the redshift and the surface mass density, as well as the intrinsic scatter of $\delta$ among the lenses, need to be properly taken into account in this method.
\end{abstract}


%%%%%%%%%%%%%%%%%%%%%%%%%%%%%%%%%%%%%%%%%%%%%%%%%%%%%%%%%%%%%%%%%%%%%%%%%%%%%%
%%%%%%%%%%%%%%%%%%%%%%%%%%%%%%%%%%%%%%%%%%%%%%%%%%%%%%%%%%%%%%%%%%%%%%%%%%%%%%
\section{Introduction}\label{sec:introduction}
 In the last two decades, owing to the advent of powerful new space and ground-based telescopes for imaging and spectroscopic observations,
 many new strong gravitational lensing (SGL) systems have been discovered. The sample size of available SGL systems has grown to be large enough for statistical analysis to study lens properties and to constrain cosmological parameters. Since the number of observed galaxy-scale SGL systems
 is much more than that of galaxy cluster-scale SGL systems, most statistical analyses have utilized the galaxy-scale SGL sample.
In practice, several different quantities can be adopted as statistical quantities with galaxy-scale SGL sample, including the distribution of image angular separations (see, e.g., Turner et al. 1984; Dyer 1984; Chiba \& Yoshii 1999; Dev et al 2004; Cao \& Zhu 2012), the distribution of lens redshifts (see, e.g., Turner et al. 1984; Kochanek 1992; Ofek et al. 2003; Mitchell et al. 2005; Cao et al. 2012a), and the velocity dispersion ($\sigma$) of lenses (see, e.g., Futamase \& Yoshida 2001; Biesiada 2006; Grillo et al. 2008; Schwab et al. 2010; Cao et al. 2017).
The major disadvantage of using the distributions of image angular separations and lens redshifts as statistical quantities is that the
theoretically predicted values of these two are dependent not only on the lens mass model but
 also on the lens luminosity function.  While the theoretical prediction of $\sigma$ is dependent only on the lens mass model but not on the luminosity function.  Besides, the gravitational lens time-delay ($\Delta\tau$) method is another cosmological application of SGL systems
  (see, e.g., Refsdal 1964; Treu \& Marshall 2016; Bonvin et al. 2017; Birrer et al. 2019), which is different
 from the three methods mentioned above, since the time-delay analysis is done for one system at a time rather than performing on a sample of lens galaxies simultaneously.
 The methods of using $\Delta\tau$ and $\sigma$ as observed  quantities for the SGL systems are both popular at present.
The theoretical analysis shows that $\Delta\tau$ is more sensitive to the cosmological
parameters than $\sigma$ (Paraficz \& Hjorth 2009; Wei \& Wu 2017). The fact also proves that the measurements of $\Delta\tau$ are very powerful at constraints on the cosmological parameters and especially sensitive to the Hubble constant $H_0$ (Bonvin et al. 2017; Suyu et al. 2017; Liao et al. 2017; Birrer et al. 2019). The measurements of $\sigma$ are weak at confining the cosmological parameters (see, e.g., Biesiada 2006; Cao et al. 2012b; Wang \& Xu 2013;  Chen et al. 2015; Cao et al. 2015;  An et al. 2016; Xia et al. 2017; Cui et al. 2017; Li et al. 2018a), but they are useful for investigating the lens mass models if the priors on cosmological parameters are given (see, e.g., Koopmans et al. 2009; Sonnenfeld et al. 2013a; Cao et al. 2016; Holanda et al. 2017).
Additionally, a combination of time delay and velocity dispersion (i.e., $\Delta\tau/\sigma^2$) is proved to be more sensitive to
 the cosmological parameters (see., e.g., Paraficz \& Hjorth 2009; Jee et al. 2015, 2016; Wei \& Wu 2017; Shajib et al. 2018) than using $\Delta\tau$ and $\sigma$ separately.

 By combining the observations of SGL and stellar dynamics in elliptical
galaxies, one can use the lens velocity dispersion (VD) as statistical quantity to put constrains on both the cosmological parameters and the density profiles
of galaxies. The core idea of this method is that the gravitational mass $M_{\textrm{grl}}^E$ and the dynamical mass $M_{\textrm{dyn}}^E$
enclosed within the disk defined by the so-called Einstein ring should be equivalent, namely, $M_{\textrm{grl}}^E=M_{\textrm{dyn}}^E$.
Further, $M_{\textrm{grl}}^E$ inferring from the strong lensing data depends on cosmological distances, and $M_{\textrm{dyn}}^E$ inferring from the stellar VD depends on both the lens mass model and the cosmological distance, so one can relate the VD with the model parameters including cosmological and lens mass model parameters.  This method can be traced back to Futamase \& Yoshida (2001), but at that time there were no available observational data of lens VD. Grillo et al. (2008) first applied this method to constrain cosmological parameters with observational data, wherein the sample included 20 SGL systems from the Lens Structure and Dynamics (LSD) survey (Koopmans \& Treu 2002, 2003; Treu \& Koopmans 2002, 2004) and the Sloan Lens ACS (SLACS) survey (Bolton et al. 2006a; Treu et al. 2006; Koopmans et al.2006).
In the literature, a recent compiled sample which can be used in this method includes 118 galaxy-scale
 SGL systems (Cao et al. 2015, hereafter C15) from the SLACS survey, the Baryon Oscillation Spectroscopic Survey (BOSS) emission-line lens survey (BELLS; see, Brownstein et al. 2012), the LSD survey,
 and the Strong Lensing Legacy Survey (SL2S; see, Gavazzi et al. 2012; Ruff et al. 2011; Sonnenfeld et al. 2013a,b, 2015). In this paper, we update the sample with definite criteria by taking advantage of new observational data, and then explore the effect of lens mass model on constraining cosmological parameters, as well as evaluate several different lens mass models.

The rest of the paper is organized as follows.  In Section 2, we demonstrate the methodology of using the lens
 velocity dispersion as statistical quantity to constrain model parameters. Then,
in Section 3 the SGL data sample used in our analysis is introduced. In Section 4, we first investigate the sensitivity of the sample under consideration to cosmological parameters, and diagnose whether the lens mass density profile is universal for the entire sample via the qualitative and semi-quantitative analysis; and then carry out observational constraints on parameters of cosmology and lens mass models.
 In the last section, the main conclusions are summarized.

\section{Methodology}\label{sec:method}
As discussed in the last section, the method of using the galaxy lens VD as statistical quantity has some special merits.
However, in this method, besides the imaging data of the SGL systems, one also has to possess the spectroscopic data
of the systems and measure the central velocity dispersion of the lens galaxies from the spectroscopy.
On the basis of various recent lensing surveys which have carried out both imaging and spectroscopic observations,
this method has become feasible.

In this method, the main idea is that the projected gravitational mass $M_{\textrm{grl}}^E$ and the projected dynamical
 mass $M_{\textrm{dyn}}^E$ within the Einstein radius should be equivalent, i.e.,
 \begin{equation}
\label{eq:Mass_eq}
M_{\textrm{grl}}^E= M_{\textrm{dyn}}^E.
\end{equation}
 From the theory of gravitational lensing,
 the projected gravitational mass within the Einstein radius is $M_{\textrm{grl}}^E=\Sigma_{cr}\pi R_E^2$. The Einstein
 radius $R_E$ is determined by $R_E=\theta_E D_l$, wherein $\theta_E$ is the Einstein angle, and $D_l$ is the angular
 diameter distance between observer and lens. The critical surface mass density $\Sigma_{cr}$ is defined
 by $\Sigma_{cr} = \frac{c^2}{4\pi G}\frac{D_s}{D_l D_{ls}}$, where $D_{ls}$ is the angular diameter distance
 between lens and source, and $D_s$ is that between observer and source. Thus, one can further figure out
\begin{equation}
\label{eq:Mgrl}
M_{\textrm{grl}}^E= \frac{c^2}{4G}\frac{D_sD_l}{D_{ls}}\theta_E^2,
\end{equation}
wherein the distances $D_s$, $D_l$ and $D_{ls}$ are dependent on the cosmological model.

To estimate the projected dynamical mass $M_{\textrm{dyn}}^E$ from the lens galaxy VD,
 one must first suppose the mass distribution model for the lens galaxy.
Here we choose a general mass model (Koopmans 2006) for the lens galaxies in our sample, which are early-type galaxies (ETGs) with E/S0 morphologies:
\begin{eqnarray}
\label{eq:profile}
\left\{
\begin{array}{lll}
\rho(r)&=& \rho_0\; (r/r_0)^{-\gamma}\\
\nu(r)&=& \nu_0\; (r/r_0)^{-\delta} \\
\beta(r)&=&1-\sigma_{\theta}^2/ \sigma_r^2
\end{array}
\right.
\end{eqnarray}
where $\rho(r)$ is the total (i.e. luminous plus dark-matter) mass density distribution, and  $\nu(r)$ is the luminosity density of stars.
The parameter $\beta(r)$ denotes the anisotropy of the stellar velocity dispersion, and is also called as the stellar orbital anisotropy, where $\sigma_\theta$ and $\sigma_r$ are
the tangential and radial velocity dispersions, respectively.

 Based on the assumption that the relationship between stellar number density $n(r)$ and stellar luminosity density $\nu(r)$ is
  spatially constant, an assumption unlikely to be violated appreciably within the effective radius of the
   early-type lens galaxies under consideration, the radial Jeans equation in Spherical Coordinate can be written as
\begin{equation}
\frac{\emph{d}}{\emph{d} r} [\nu(r)\sigma^2_r]+\frac{2\beta}{r}\nu(r)\sigma^2_r =-\nu(r)\frac{\emph{d} \Phi}{\emph{d} r},
\label{eq:Radial_JeansEq}
\end{equation}
where
 \begin{equation}
\frac{\emph{d}\Phi}{\emph{d}r}=\frac{GM(r)}{r^2},
\label{eq:dPhi_dr}
\end{equation}
and $M(r)$ is the total mass inside a sphere with radius $r$.
By substituting Eq. (\ref{eq:dPhi_dr}) into Eq. (\ref{eq:Radial_JeansEq}), one can get the expression for $\sigma_r^2$,
\begin{equation}
\sigma_r^2(r) = \frac{G\int^{\infty}_{r}\emph{d}r^{\prime}r^{\prime 2\beta-2}\nu(r^{\prime})M(r^{\prime})}{r^{2\beta}\nu(r)},
\label{eq:sigma2r}
\end{equation}

By defining $r$ to be the spherical radial coordinate from the lens center, $Z$ to be the axis along the line of sight (LOS) , and $R$ to be the cylindrical radius which is perpendicular to the LOS, then one has $r^2=R^2+Z^2$. The projected dynamical
 mass $M_{\textrm{dyn}}$  contained within a cylinder of radius equal to the Einstein radius $R_E$ can be calculated with
\begin{equation}
M_{\textrm{dyn}}^E = \int_{0}^{R_E}\emph{d}R 2\pi R'\Sigma(R'),
\label{eq:ME}
\end{equation}
where
\begin{eqnarray}
\Sigma(R)&=&\int_{-\infty}^{\infty}\rho(r)dZ\nonumber\\
&=& \int_{-\infty}^{\infty}\emph{d}Z \frac{\rho_0}{r_0^{-\gamma}}(Z^2+R^2)^{-\gamma/2}\nonumber\\
&=&\sqrt(\pi)R^{1-\gamma}\frac{\Gamma\left(\frac{\gamma-1}{2}\right)}{\Gamma(\gamma/2)}\frac{\rho_0}{r_0^{-\gamma}}
\label{eq:Sigma_R}
\end{eqnarray}
By substituting Eq.(\ref{eq:Sigma_R}) into Eq.(\ref{eq:ME}), one can have
 \begin{equation}
M_{\textrm{dyn}}^E = 2\pi^{3/2}\frac{R_E^{3-\gamma}}{3-\gamma}\frac{\Gamma\left(\frac{\gamma-1}{2}\right)}{\Gamma(\gamma/2)}\frac{\rho_0}{r_0^{-\gamma}}.
\label{eq:ME2}
\end{equation}
The total mass contained within a sphere with radius $r$ is
\begin{equation}
M(r)=\int_0^r dr' 4\pi r'^2\rho(r')=4\pi\frac{\rho_0}{r_0^{-\gamma}}\frac{r^{3-\gamma}}{3-\gamma}.
\label{eq:Mr}
\end{equation}
By combining Eqs. (\ref{eq:ME2}) and (\ref{eq:Mr}), one can further have
\begin{equation}
M(r)=\frac{2}{\sqrt{\pi}}\frac{\Gamma(\gamma/2)}{\Gamma(\frac{\gamma-1}{2})}\left(\frac{r}{R_E}\right)^{3-\gamma}M_{\textrm{dyn}}^E.
\label{eq:Mr2}
\end{equation}
By substituting Eqs. (\ref{eq:Mr2}) and (\ref{eq:profile}) into Eq. (\ref{eq:sigma2r}), one reads
\begin{equation}
\sigma_r^2(r)
=
\frac{2}{\sqrt{\pi}}\frac{GM_{\textrm{dyn}}^E}{R_E}\frac{1}{\xi-2\beta}\frac{\Gamma(\gamma/2)}
{\Gamma(\frac{\gamma-1}{2})}\left(\frac{r}{R_E}\right)^{2-\gamma},
\label{eq:sigma_2r_2}
\end{equation}
where $\xi=\gamma+\delta-2$, and $\beta$ is assumed to be independent of the radius $r$.

The actual velocity dispersion of the lens galaxy measured by the observation is the component of luminosity-weighted average along the LOS and over the effective spectroscopic aperture $R_A$, that can be expressed mathematically
\begin{equation}
\sigma^2_{\parallel}(\leq R_A) = \frac{\int_{0}^{R_A}\emph{d}R\, 2\pi R\int_{-\infty}^{\infty}\emph{d}Z\, \sigma^2_{\textrm{los}}\nu(r)}{\int_{0}^{R_A}\emph{d}R\, 2\pi R\int_{-\infty}^{\infty}\emph{d}Z\,\nu(r)}
\label{eq:sigma2ll_1}
\end{equation}
where $\sigma^2_{\textrm{los}}$ is the LOS velocity dispersion, which is a combination of the radial ($\sigma_r^2$) and tangential ($\sigma_t^2$) velocity dispersions. Using $\theta$ to indicate the angle between the LOS (Z-axis) and the radial direction (r-axis), then
one reads
\begin{eqnarray}
\sigma^2_{los}&=&(\sigma_r \cos\theta)^2+(\sigma_t \sin\theta)^2\nonumber\\
&=&\sigma_r^2\frac{r^2-R^2}{r^2}+\sigma_t^2\frac{R^2}{r^2}\nonumber\\
&=&\sigma_r^2(1-\frac{R^2}{r^2})+(1-\beta)\sigma_r^2\frac{R^2}{r^2}\nonumber\\
&=&\sigma_r^2(1-\beta\frac{R^2}{r^2})
\label{eq:sigma_los}
\end{eqnarray}
By substituting Eq.(\ref{eq:sigma_los}) into Eq.(\ref{eq:sigma2ll_1}), one can read
\begin{equation}
\sigma^2_{\parallel}(\leq R_A) = \frac{\int_{0}^{R_A}\emph{d}R\, 2\pi R\int_{-\infty}^{\infty}\emph{d}Z\, \sigma^2_r(r)(1-\beta\frac{R^2}{r^2})\nu(r)}{\int_{0}^{R_A}\emph{d}R\, 2\pi R\int_{-\infty}^{\infty}\emph{d}Z\,\nu(r)}
\label{eq:sigma2ll_2}
\end{equation}
Further, by substituting Eq.(\ref{eq:sigma_2r_2}) and (\ref{eq:profile}) into Eq.(\ref{eq:sigma2ll_2}), one obtains
\begin{eqnarray}
\begin{array}{rr}
\sigma^2_{\parallel}(\leq R_A) = \frac{2}{\sqrt{\pi}}\frac{GM_{\textrm{dyn}}^E}{R_E}\frac{3-\delta}{(\xi-2\beta)(3-\xi)}\left[\frac{\Gamma
\left[(\xi-1)/2\right]}{\Gamma(\xi/2)}-\beta\frac{\Gamma\left[(\xi+1)/2\right]}{\Gamma\left[(\xi+2)/2
\right]}\right]\\
\frac{\Gamma(\gamma/2)\Gamma(\delta/2)}{\Gamma\left[(\gamma-1)/2\right]\Gamma\left[(\delta-1)/2\right]}
\left(\frac{R_A}{R_E}\right)^{2-\gamma}.
\end{array}
\end{eqnarray}
Finally, with the relation expressed in Eq.(\ref{eq:Mass_eq}), the above formula can be rewritten as
\begin{eqnarray}
\label{eq:sigma_thetaA}
\begin{array}{rr}
\sigma^2_{\parallel}(\leq R_A) = \frac{c^2}{2\sqrt{\pi}}\frac{D_s}{D_{ls}}\theta_E\frac{3-\delta}{(\xi-2\beta)(3-\xi)}\left[\frac{\Gamma
\left[(\xi-1)/2\right]}{\Gamma(\xi/2)}-\beta\frac{\Gamma\left[(\xi+1)/2\right]}{\Gamma\left[(\xi+2)/2
\right]}\right]\\
\frac{\Gamma(\gamma/2)\Gamma(\delta/2)}{\Gamma\left[(\gamma-1)/2\right]\Gamma\left[(\delta-1)/2\right]}
\left(\frac{\theta_A}{\theta_E}\right)^{2-\gamma},
\end{array}
\end{eqnarray}
where $R_A=\theta_A D_l$.

 From the spectroscopic data, one can measure the velocity dispersion $\sigma_{\textrm{ap}}$ inside the circular aperture with
 the angular radius $\theta_{\textrm{ap}}$. In practice, if the $\sigma_{\textrm{ap}}$ are measured within rectangular apertures,
one usually derives the equivalent circular apertures with
the angular radii $\theta_{\textrm{ap}}$ following J{\o}rgensen et al. (1995),
\begin{equation}
\label{eq:theta_ap_eff}
\theta_{\textrm{\textsf{ap}}}\approx 1.025\times \sqrt{(\theta_x\theta_y/\pi)},
\end{equation}
where $\theta_x$ and $\theta_y$ are the angular sizes of
width and length of the rectangular aperture.
 More precisely, $\sigma_{\textrm{ap}}$ is the luminosity weighted average of the line-of-sight velocity dispersion of the lensing galaxy inside $\theta_{\textrm{ap}}$.
For a fair comparison and in consideration of the effect of the aperture size on the measurements of velocity dispersions, all
velocity dispersions $\sigma_{\textrm{ap}}$  measured within apertures of arbitrary sizes,
are normalized to a typical physical aperture, $\sigma_{\textrm{e2}}$, with the
 radius $R_{\textrm{eff}}/2$, where $R_{\textrm{eff}}$ is the half-light radius of the lens galaxy. The radius $R_{\textrm{eff}}/2$ is chosen because it is well-matched to the typical
Einstein radius, therefore just a small error is brought in when the relation satisfied in the Einstein radius (e.g., Eq.(\ref{eq:Mass_eq}) ) is extrapolated to the radius $R_{\textrm{eff}}/2$ (Auger et al. 2010).
Following the prescription, one can use the aperture correction formula,
\begin{eqnarray}
\label{eq:sigma_obs}
\sigma^{\textrm{obs}}_{\parallel} \equiv \sigma_{\textrm{e2}} = \sigma_{\textrm{ap}}[\theta_{\textrm{eff}}/(2\theta_{\textrm{ap}})]^{\eta},
\end{eqnarray}
where $\theta_{\textrm{eff}} = R_{\textrm{eff}}/D_l$.
The best-fitting values of the
correction factor $\eta$ are different when using different observational samples. For example, the best-fitting values of $\eta$ are $-0.04$, $-0.06$ and $-0.066\pm 0.035$ found by J{\o}rgensen et al. (1995), Mehlert et al. (2003) and Cappellari et al. (2006), respectively, where the third value is consistent with the former two at the 1$\sigma$ level.
 In this work, we adopt the value $\eta = -0.066\pm 0.035$ from Cappellari et al. (2006). Then, the total uncertainty of $\sigma_{\textrm{e2}}$, i.e., $\Delta\sigma_{\textrm{e2}}^{\textrm{tot}}$,  satisfies
\begin{equation}
\label{eq:err_sigma_e2}
 (\Delta\sigma_{\textrm{e2}}^{\textrm{tot}})^2 = (\Delta\sigma_{\textrm{e2}}^{\textrm{stat}})^2+(\Delta\sigma_{\textrm{e2}}^{\textrm{AC}})^2+
 (\Delta\sigma_{\textrm{e2}}^{\textrm{sys}})^2.
\end{equation}
The the statistical error, $\Delta\sigma_{\textrm{e2}}^{\textrm{stat}}$, is propagated from the measurement error of $\sigma_{\textrm{ap}}$. The error due to the aperture correction, $\Delta\sigma_{\textrm{e2}}^{\textrm{AC}}$, is propagated from the uncertainty of $\eta$.
In addition to the measurement errors, we should also consider the systematic error $\Delta\sigma_{\textrm{e2}}^{\textrm{sys}}$. The essential assumption of the method is that the projected mass within the Einstein radius, $M^E$, can be uniformly estimated from both the gravitational and dynamical masses, i.e.,  $M^E = M^E_\textrm{grl} = M^E_\textrm{dyn} $. In practice, the model-predicted value of $M^E = M^E_
\textrm{dyn}$ from Eq.(\ref{eq:ME2}) only includes the contribution from the lens galaxy, while the value of $M^E = M^E_\textrm{grl}$ from Eq.(\ref{eq:Mgrl}) includes the extra contribution from other matters (outside of the lens galaxy) along the line of sight. The extra mass from the lensing data
 can be treated as a systematic error, which contributes uncertainty of $\sim$3\%  to the model-predicted value of the velocity dispersion (Jiang \& Kochanek 2007).

In order to compare the observational values of the VD with the corresponding model-predicted ones, one needs to
calculate the theoretical value of the VD within the radius $R_{\textrm{eff}}/2$ from Eq. (\ref{eq:sigma_thetaA}) (Koopmans 2006),
\begin{equation}
\label{eq:sigma_th}
\sigma_{\parallel(\leq\theta_{\textrm{eff}}/2)}=\sqrt{\frac{c^2}{2\sqrt{\pi}}\frac{D_s}{D_{ls}}\theta_E\frac{3-\delta}{(\xi-2\beta)(3-\xi)}F(\gamma,\delta, \beta)\left(\frac{\theta_{\textrm{eff}}}{2 \theta_\textrm{E}}\right)^{(2-\gamma)}},
\end{equation}
where
\begin{equation}
F=\left[\frac{\Gamma
\left[(\xi-1)/2\right]}{\Gamma(\xi/2)}-\beta\frac{\Gamma\left[(\xi+1)/2\right]}{\Gamma\left[(\xi+2)/2
\right]}\right]
\frac{\Gamma(\gamma/2)\Gamma(\delta/2)}{\Gamma\left[(\gamma-1)/2\right]\Gamma\left[(\delta-1)/2\right]}.
\end{equation}
In the case of $\gamma = \delta =2$ and $\beta= 0$, the mass model is reduced to the well-known Singular Isothermal Sphere(SIS) model,
and the predicted value of the VD is recovered to
\begin{equation}
\label{eq:sigma_th_SIS}
\sigma_{\textrm{SIS}}= \sqrt{\frac{c^2}{4\pi}\frac{D_{s}}{D_{ls}}\theta_E}.
\end{equation}

In our analysis, the likelihood is assumed to be
\begin{equation}
\label{eq:likelihood}
\mathcal{L}\propto e^{-\chi^2/2}.
\end{equation}
$\chi^2$ is constructed as
\begin{equation}
\label{eq:chi2}
\chi^2=\sum^{N}_{i=1}\left(\frac{\sigma^{\textrm{th}}_{\parallel ,i}-\sigma^{\textrm{obs}}_{\parallel ,i}}
{\Delta\sigma^{\textrm{tot}}_{\parallel ,i}}\right)^2,
\end{equation}
where $N$ is the number of the data points, $\Delta\sigma^{\textrm{tot}}_{\parallel ,i}$ is
the uncertainty of $\sigma^{\textrm{obs}}_{\parallel ,i}$, which is calculated with Eq. (\ref{eq:err_sigma_e2}). One can obtain $\sigma^{\textrm{obs}}_{\parallel ,i}$ and $\sigma^{\textrm{th}}_{\parallel ,i}$
from Eqs. (\ref{eq:sigma_obs}) and (\ref{eq:sigma_th}), respectively.

In the following analyses, we derive the posterior probability distributions of model parameters through an affine--invariant Markov chain Monte Carlo (MCMC) Ensemble sampler (emcee; Foreman-Mackey et al. 2013), where the likelihood is computed with Eqs. (\ref{eq:likelihood}) and (\ref{eq:chi2}).
For the purpose of the analysis in this work, it would suffice to assume that $\beta$ is independent of $r$ (see, e.g., koopmans et al. 2006; Treu et al. 2010). Because we cannot
independently measure $\beta$ for individual lensing systems, We then treat $\beta$
as a nuisance parameter and marginalize over it using a Gaussian prior with  $\beta=0.18\pm0.13$, that is an independent constraint on $\beta$ from a well-studied sample of nearby elliptical galaxies
(see, e.g.,Gerhard et al. 2001), and adopted in the previous works (see, e.g., Bolton et al. 2006; Schwab et al. 2010; Cao et al. 2017).  Thus, throughout this paper a Gaussian prior on $\beta$ with $\beta=0.18\pm0.13$ is used over the range of $[\bar{\beta}-2\sigma_{\beta}, \bar{\beta}+2\sigma_{\beta}]$ where $\bar{\beta}=0.18$ and $\sigma_{\beta}=0.13$, unless some special instructions are made.
 In addition, we assume a flat prior for each remaining parameter over a range of interest.

\section{Data sample}\label{sec:data}
According to the analysis in the last section, one can learn that the method under consideration requires the following information
from observations, including the lens redshift $z_l$, the source redshift $z_s$, the Einstein angle $\theta_E$, the central VD of
the lens galaxy $\sigma_{\textrm{ap}}$, the spectroscopic aperture angular radius $\theta_{\textrm{ap}}$, and the half-light angular radius of the
lens galaxy $\theta_{\textrm{eff}}$. Additionally,  to ensure the validity of the assumption of spherical symmetry on the lens galaxy,
the selected lens galaxies should satisfy the following
conditions:(i) the lens galaxy should be ETGs with E/S0 morphologies; and (ii) the lens galaxy should
not have significant substructure or close massive companion.
Some lens galaxies from C15's sample do not satisfy the
above conditions. Here we assemble a sample including 161 galaxy-scale SGL systems which meet all the requirements mentioned above,
 where 5 systems from the LSD survey\footnote{http://web.physics.ucsb.edu/$\sim$tt/LSD/}(Koopmans \& Treu 2002, 2003; Treu \& Koopmans 2002, 2004), 26 from the SL2S (Ruff et al. 2011; Sonnenfeld et al. 2013a,b; Sonnenfeld et al. 2015), 57 from the SLACS (Bolton et al. 2008; Auger et al. 2009, 2010), 38 from the an extension of the SLACS survey known
as ``SLACS for the Masses'' (hereafter S4TM, Shu et al. 2015; Shu et al. 2017), 21 from the BELLS (Brownstein et al. 2012, hereafter B12), and 14 from
the BELLS for GALaxy-Ly$\alpha$ EmitteR sYstemsGALLERY (hereafter BELLS GALLERY, Shu et al. 2016a,b). The useful information of these 161 systems is listed in Appendix (i.e., Table \ref{tab:sample}). The SLACS lenses used in this work are selected from the full SLACS sample (Bolton et al. 2008, hereafter B08) with high-fidelity observations carried out using the Advanced Camera for Surveys (ACS) on Hubble Space Telescope (HST), where the data of $\theta_{E}$ and $\theta_{\textrm{eff}}$ are taken from Tables 4 and 5 of B08, and the data of $z_l$, $z_s$ and $\sigma_{\textrm{ap}}$ from the Table 3 of Auger et al. (2009). All the observational data of BELLS lenses are taken from B12. In B12, the effective radii $\theta_{\textrm{eff}}$ of the lenses are measured from both the BOSS and HST-ACS imaging data, and the measurements from the later are much more precise than those from the former. So, We choose to use the $\theta_{\textrm{eff}}$ data from the HST-ACS observations, which are listed in Table 3 of B12. In addition, the lenses from S4TM and  BELLS GALLERY surveys are not included in C15's sample.

\begin{figure*}
\centering
\includegraphics[width=0.48\linewidth]{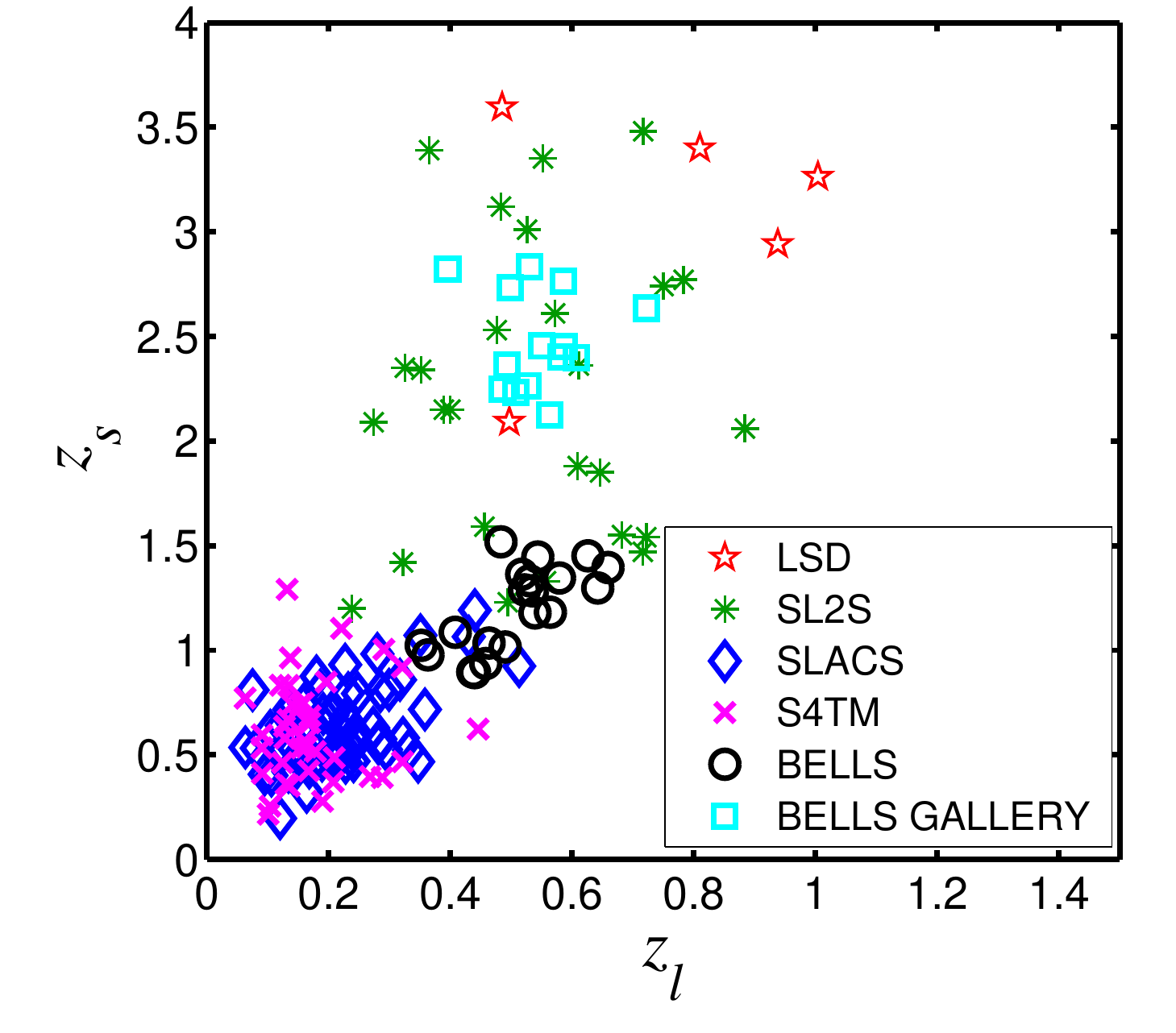}
\includegraphics[width=0.48\linewidth]{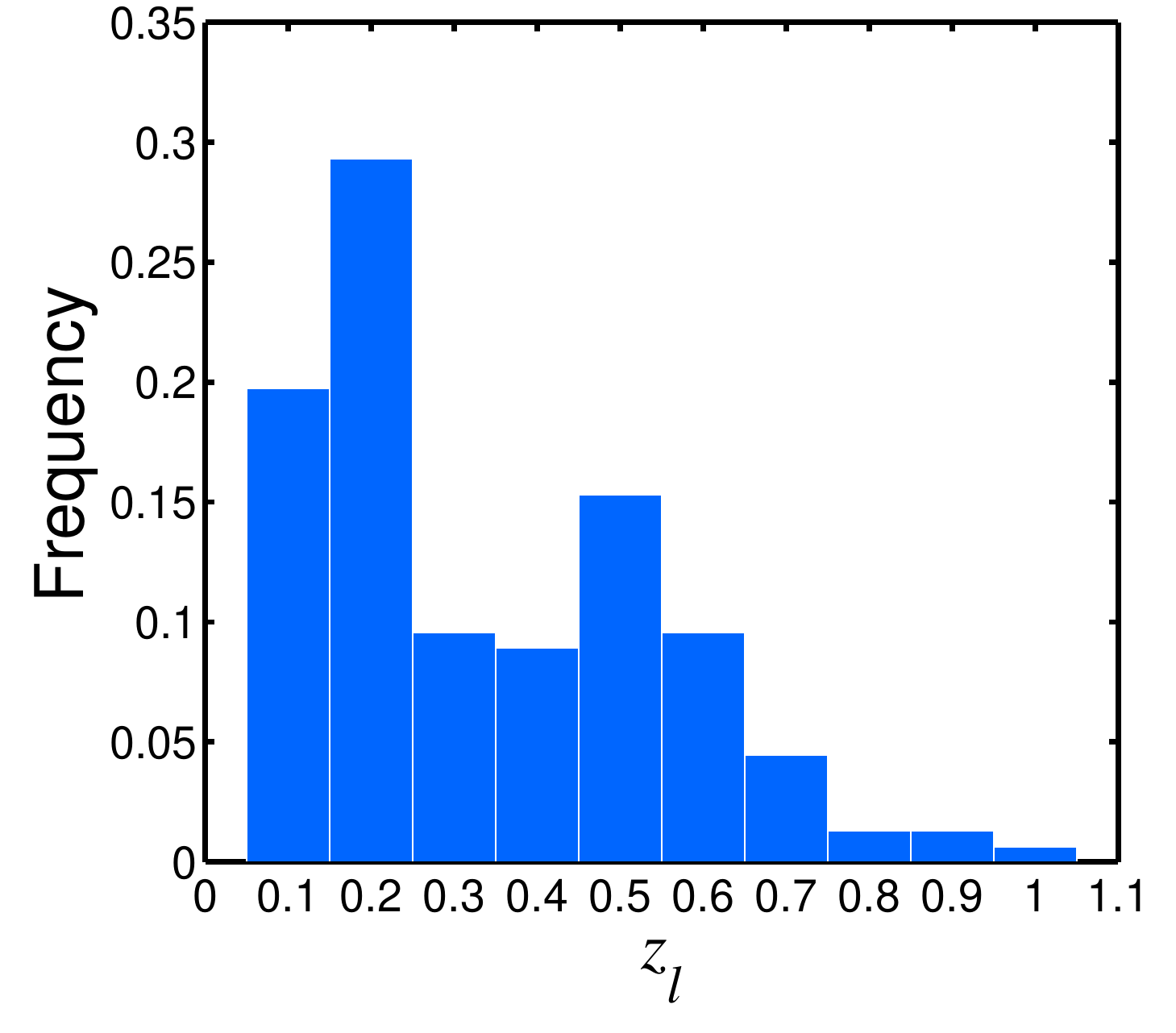}\\
\includegraphics[width=0.48\linewidth]{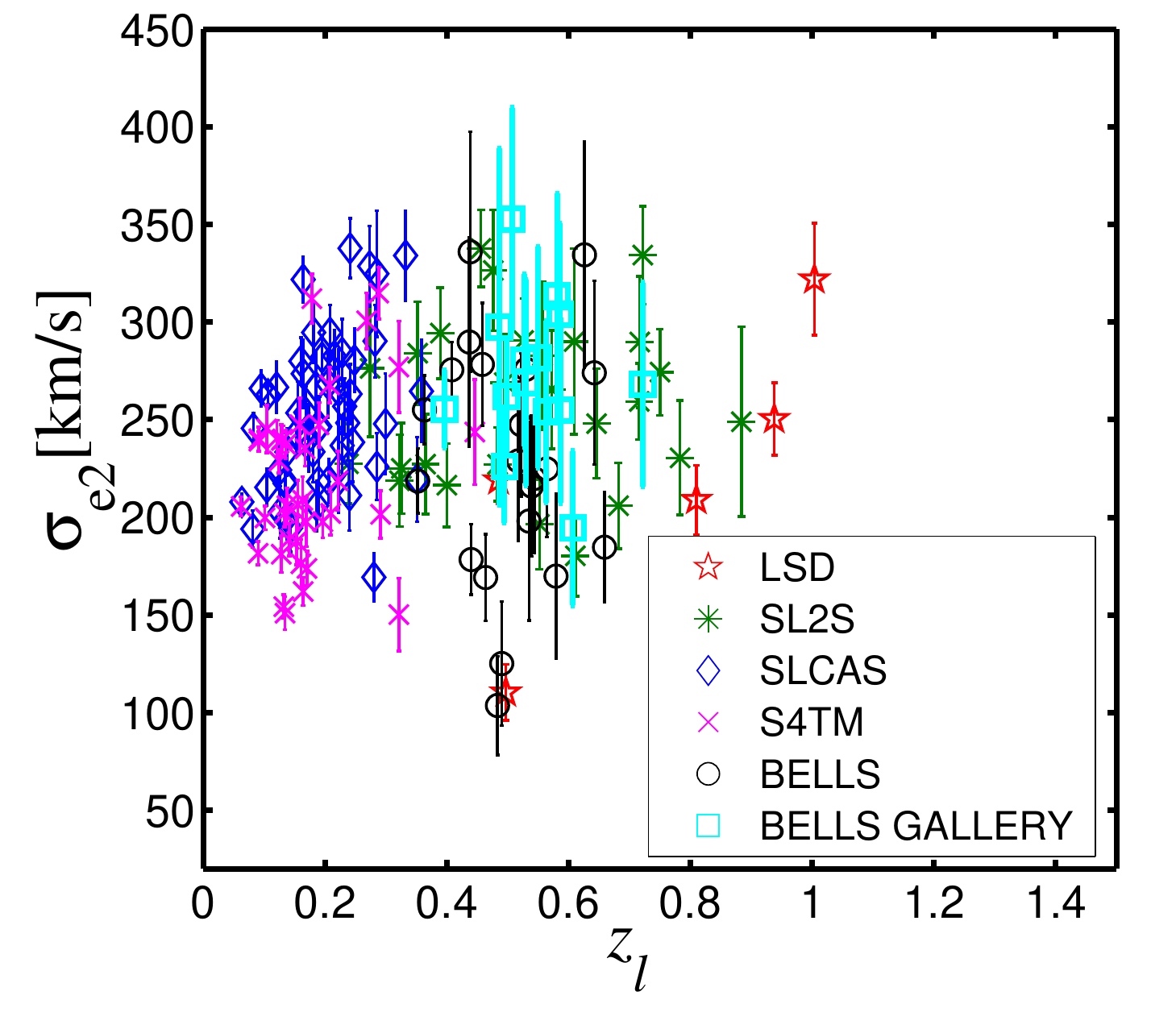}
\includegraphics[width=0.48\linewidth]{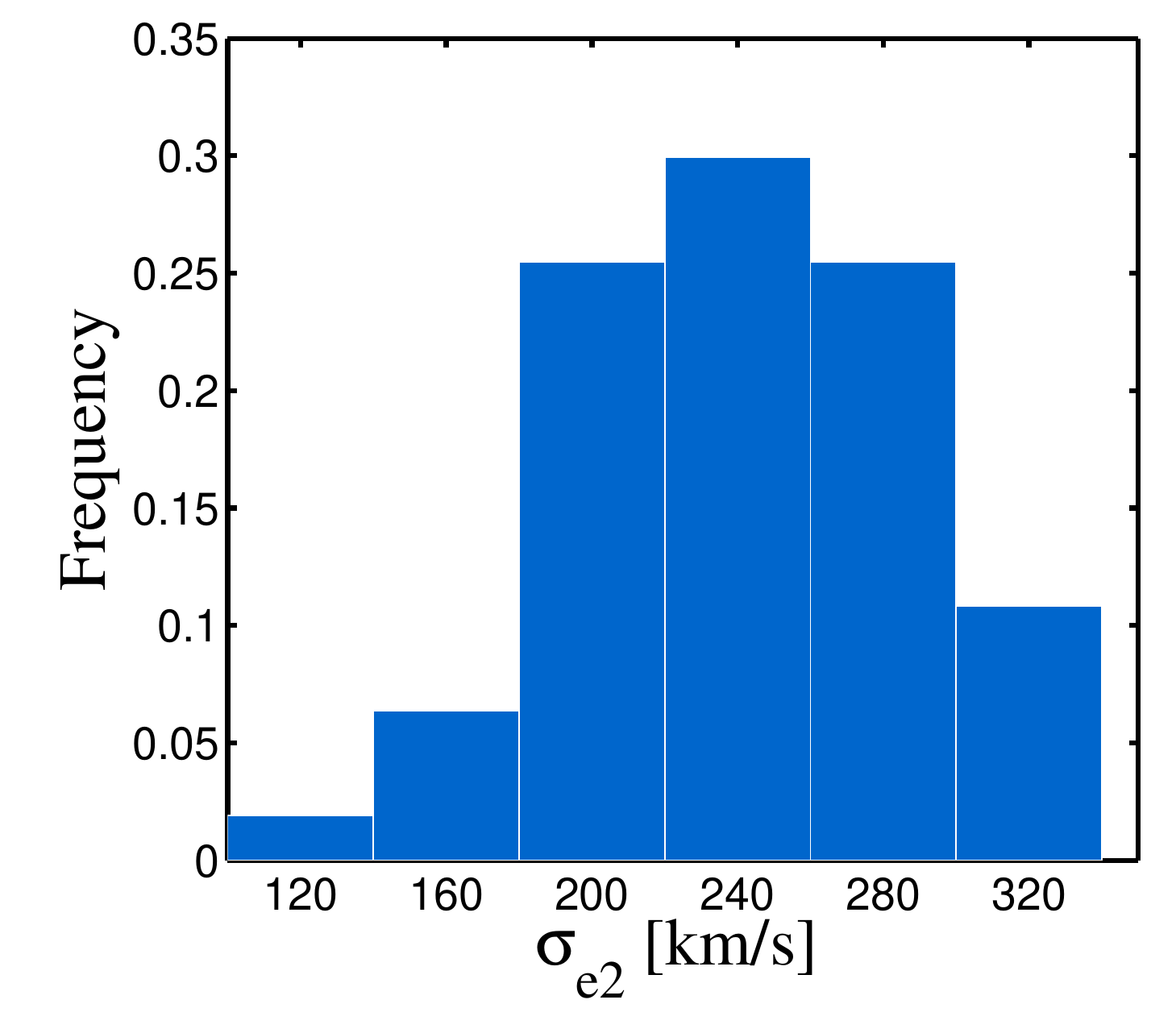}
 \caption{The distributions of the lens redshifts $z_l$, the source redshifts $z_s$ and the lens stellar velocity dispersions $\sigma_{e2}$ for the SGL sample presented in Table \ref{tab:sample}. The left panels show the sample distribution in $z_l$ -- $z_s$ and $z_l$ -- $\sigma_{e2}$ planes, where the points with different colors denote samples from different surveys. The right panels display the normalized histograms of $z_l$ and $\sigma_{e2}$ for the whole sample. We note that the shown error bars are the total errors ($\Delta \sigma^{\textrm{tot}}_{\textrm{e2}}$) calculated with Eq. (\ref{eq:err_sigma_e2}) in the  $z_l$ -- $\sigma_{e2}$ plane.}
\label{fig:distributions}
\end{figure*}

The velocity dispersions of the lenses from LSD and SL2S surveys, which are measured within rectangular slits, are transformed into velocity dispersions  $\sigma_{e2}$, within a circular aperture with radius $R_{\textrm{eff}}/2$ based on Eqs.(\ref{eq:theta_ap_eff}) and (\ref{eq:sigma_obs}). The SLACS and S4TM surveys select candidates from Sloan Digital Sky Survey I (SDSS-I, Eisenstein et al. 2001; Strauss et al. 2002) data, in which the velocity dispersions of the lenses are measured within the
$1.5^{\prime\prime}$--radius fibers. The lens candidates of the BELLS and BELLS GALLERY surveys are spectroscopically selected from the BOSS (Dawson et al. 2013) of the Sloan Digital Sky Survey-III (SDSS-III, Eisenstein et al.
2011), in which the VD of the lenses are measured within the
$1^{\prime\prime}$--radius fibers. These velocity dispersions measured with fibers are corrected to $\sigma_{e2}$ based on Eq.(\ref{eq:sigma_obs}). The distribution of the whole SGL sample is shown in Figure \ref{fig:distributions}.
From the upper panels of Figure \ref{fig:distributions}, one can see that $\sim$30\% of the lenses  are located at $z_l \sim 0.2$, and only $\sim$5\% located at $z_l>0.75$. The lower panels of Figure \ref{fig:distributions} show that  $\sim$80\% of the lenses possess the velocity dispersions $180\, \textrm{km}\,\textrm{s}^{-1} <\sigma_{e2}<300\,\textrm{km}\,\textrm{s}^{-1}$.

\section{Analysis and Results}\label{sec:analysis}
\subsection{Qualitative analysis}

\begin{figure*}
\centering
\includegraphics[width=0.98\linewidth]{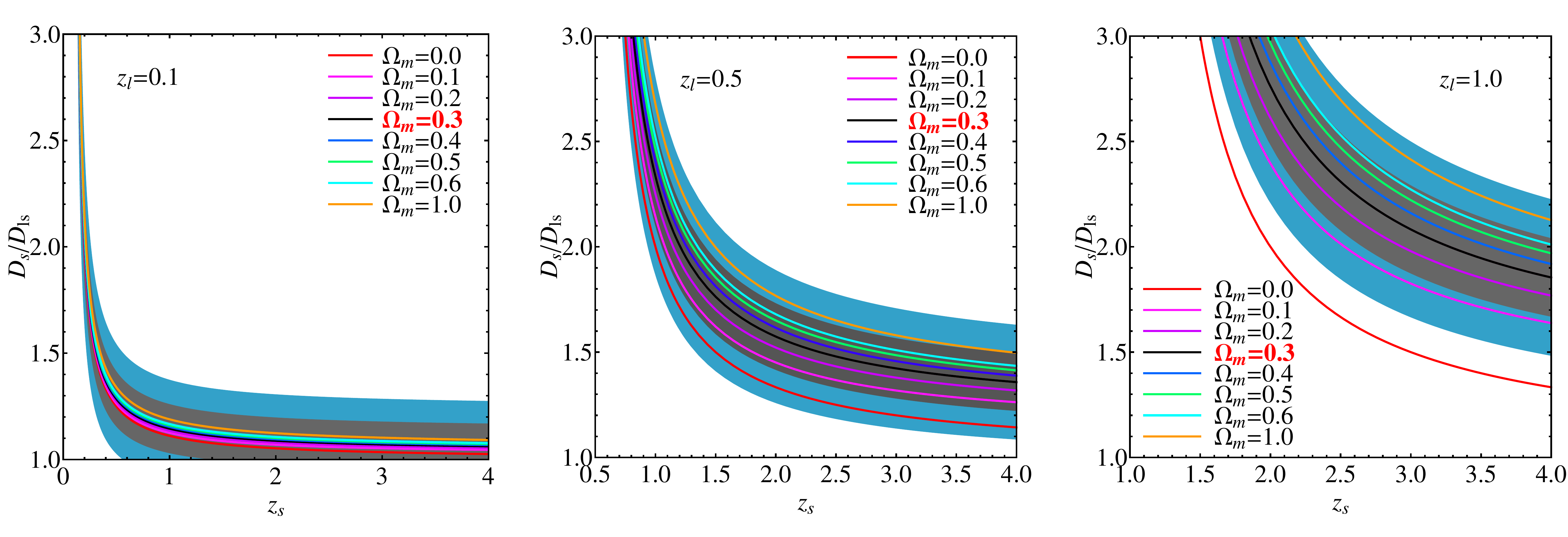}
 \caption{The evolution of the distance ratio $D_s/D_{ls}$  with respect to source redshift $z_s$ with the variety of $\Omega_m$, corresponding to lens redshift $z_l = 0.1, 0.5\; \textrm{and}\;1$, where the other cosmological parameters take the fiducial
values. In each panel, the solid black line corresponds to the fiducial value of $D_s/D_{ls}$, which is plotted with the fiducial values of cosmological parameters, and the shadows denote the cases that the relative uncertainties of
$D_s/D_{ls}$ are $10\%$ and $20\%$ with respect to the fiducial value.}
\label{fig:sen_om}
\end{figure*}

From Eq. (\ref{eq:sigma_th}) one can see that the cosmological model enters into the theoretical observable $\sigma^{\textrm{th}}_{\parallel}$  not through a distance measure directly, but rather through a distance ratio
\begin{equation}
\label{eq:Dth}
\frac{D_s}{D_{ls}}=\frac{\int_0^{z_s} \frac{dz}{E(z;\textbf{p})}}{\int_{z_l}^{z_s} \frac{dz}{E(z;\textbf{p})}},
\end{equation}
where in the framework of the flat FLRW metric the theoretical values of $D_s$ and $D_{ls}$ can be obtained by
\begin{equation}
\label{eq:Dsth} D_s(z_s; \textbf{p}, H_0) = \frac{c}{H_0(1+z_s)} \int_0^{z_s} \frac{dz}{E(z;\textbf{p})},
\end{equation}
and
\begin{equation}
\label{eq:Ddsth} D_{ls}(z_l, z_s; \textbf{p}, H_0) = \frac{c}{H_0(1+z_s)} \int_{z_l}^{z_s} \frac{dz}{E(z;\textbf{p})},
\end{equation}
respectively, where  $\textbf{p}$ denotes the parameter space of the considered cosmological model, and $E=H/H_0$ is the dimensionless Hubble parameter, and $c$ is the velocity of light. The theoretical prediction of the  observable is independent of the Hubble constant $H_0$
which gets canceled in the distance ratio. On the other side, the distance ratio $D_s/D_{ls}$ is a ratio of two integrals which have the same integrand (i.e., $1/E(z;\textbf{p})$) and differ only by the limits of integration, so the theoretical observable $\sigma^{\textrm{th}}_{\parallel}\propto \sqrt{D_s/D_{ls}}$ is insensitive to the cosmological parameters involved in the integrand (Biesiada et al. 2010). In Figure \ref{fig:sen_om}, we show the impact of the matter-density parameter $\Omega_m$ on the distance ratio by taking a spatially flat $\Lambda$CDM model with $\Omega_m = 0.3$ as a fiducial cosmological model. The three panels of Figure \ref{fig:sen_om} display the evolution of $D_s/D_{ls}$ with respect to the source redshift $z_s$ along with variety of $\Omega_m$, corresponding to the cases of the lens redshift $z_l = 0.1, 0.5, \textrm{and} 1$ from left to right. The general trend is that the sensitivity of $D_s/D_{ls}$ to $\Omega_m$ increases with the increase of $z_l$.
In Figure \ref{fig:sen_om}, the shadows denote the cases that the relative uncertainties of $D_s/D_{ls}$ are 10\% and 20\%, respectively, with
respect to the fiducial value. One can see that an individual lens with $z_l = 0.1$ cannot put any constraint on $\Omega_m$
even when $D_s/D_{ls}$ only has 10\% uncertainty.  An individual lens with $z_l = 0.5$ can bound on $\Omega_m$ with
$\sim$80\%--160\% relative uncertainty when $D_s/D_{ls}$ only has 10\% uncertainty, but cannot put any constraint on $\Omega_m$
when the uncertainty of $D_s/D_{ls}$ increases to 20\%.  Unfortunately, with regard to the SGL sample under consideration,
the typical values of the relative uncertainties of $D_s/D_{ls}$
 are approximately 10\% and 20\% at $z_l \approx 0.1$ and $0.5$, respectively\footnote{The uncertainty on $D_s/D_{ls}$ is mainly propagated from that on $\sigma_{e2}$. The relative uncertainty on
$D_s/D_{ls}$ is about 2 times of that on $\sigma_{e2}$ because of $D_s/D_{ls}\propto \sigma_{e2}^2$.}. It means that most individual lenses with $z_l < 0.5$ in our sample do not contribute to the limit on $\Omega_m$. An
individual lens with $z_l = 1$ can put a limit on $\Omega_m$ with $\sim$ 50\% -- 100\% ($\sim$ 80\% -- 200\%) relative uncertainty,
corresponding to $D_s/D_{ls}$ with 10\%(20\%) uncertainty. In our sample, there is only one system with $z_l > 1$, that is MG2016$+$112 with
$z_l = 1.004$ from LSD survey. In general, one is not able to make a high-precision estimate on $\Omega_m$ with the sample under consideration.
After repeating similar analyses for other cosmological parameters (i.e., the equation of state parameter of dark energy, and the curvature parameter), we find out that the current sample is really weak at confining these cosmological parameters. Besides, the distance ratio $D_s/D_{ls}$ is more sensitive to $\Omega_m$ than to the equation of state parameter of dark energy (Sereno 2002).

\subsection{Observational constraints} \label{sec:Obs_Cons}

We assume a kind of spherically symmetric mass distributions (i.e. Eq. (\ref{eq:profile})) for the lens galaxies in the kinematic analysis.  As discussed above, the dependence of $\gamma$ on the properties of lens galaxies should be taken into account. In the previous works, the dependence of the total mass density slope $\gamma$ on the redshift has been widely studied (see, e.g., Ruff et al. 2011; Bolton et al 2012; Cao et al. 2015; Cao et al. 2016; Cui et al. 2017; Holanda et al. 2017). Besides, Auger et al. (2010) also found a significant correlation between $\gamma$ and total mass surface density, that has also been confirmed by Dutton \& Treu (2014) and Sonnenfeld et al. (2013a). In the light of these works, we specifically consider three parameterizations for $\gamma$, namely:
\begin{itemize}
\item $P_1:$   $\gamma =\gamma_0,$
\item $P_2:$  $\gamma=\gamma_0+\gamma_z*z_l,$
\item $P_3:$ $\gamma=\gamma_0+\gamma_z*z_l+\gamma_s*\log\tilde{\Sigma},$
\end{itemize}
where $\gamma$ is treated as an arbitrary constant in case $P_1$, and its dependence on the lens redshift $z_l$ is considered in
case $P_2$.  Besides, the dependence on both the redshift and the surface mass density is taken into account in case $P_3$. According to the virial theorem, the projected dynamical mass within the radius $R_{\textrm{eff}}/2$ satisfies $M_{\textrm{e2}}^{\textrm{dyn}}\propto\sigma_{\textrm{e2}}^2R_{\textrm{eff}}$ (see, e.g., Auger et al. 2010), so the corresponding surface mass density is $\Sigma \propto \sigma_{\textrm{e2}}^2/R_{\textrm{eff}}$.
Here, we use $\tilde{\Sigma}$ to denote the normalized surface mass density of the lens galaxy, which is expressed as
\begin{equation}
\tilde{\Sigma}=\frac{(\sigma_{\textrm{e2}}/100 \textrm{km}\, \textrm{s}^{-1})^2}{R_{\textrm{eff}}/10h^{-1}\textrm{kpc}},
\end{equation}
where the usual convention of writing the Hubble constant as $H_0 = 100 h$ $\textrm{km} \textrm{s}^{-1} \textrm{Mpc}^{-1}$ is adopted.

As mentioned above, the sample under consideration is quite weak at constraining cosmological parameters, so constraining too many cosmological parameters simultaneously would only distort the results. Thus, we only attempt to fit $\Omega_m$ in the framework of flat $\Lambda$CDM model, where  $\Omega_m$ is the only free parameter of cosmology. We then conduct observational constraints on $\Omega_m$ and lens mass model parameters in the scenarios of ``$P_1$'', ``$P_2$'' and ``$P_3$'', respectively. As discussed at the end of Sec. {\ref{sec:method}}, the orbit anisotropy
parameter $\beta$ is treated as a nuisance parameter and marginalize over it using a Gaussian
prior, and the flat priors are assumed for other free parameters.  In the following, we consider two different treatment schemes for the slope ($\delta$) of
the luminosity density profile.

\subsubsection{The case of treating $\delta$ as a universal parameter for all lens galaxies in the sample}

We first consider the case of treating the luminosity density slope $\delta$  as a universal parameter for all lens galaxies in the entire sample. In other words, the intrinsic scatter of $\delta$ among the lens galaxies is assumed to be ignorable.
This treatment scheme to $\delta$ is the same as that adopted in  Cao et al. (2016, 2017) and Xia et al. (2017).
In the previous studies, another treatment to $\delta$ is setting $\delta = \gamma$ (see., e.g., C15; An et al. 2016; Cui et al. 2017),  that is
not adopted in this work.

\begin{figure*}
\centering
\includegraphics[width=0.95\linewidth]{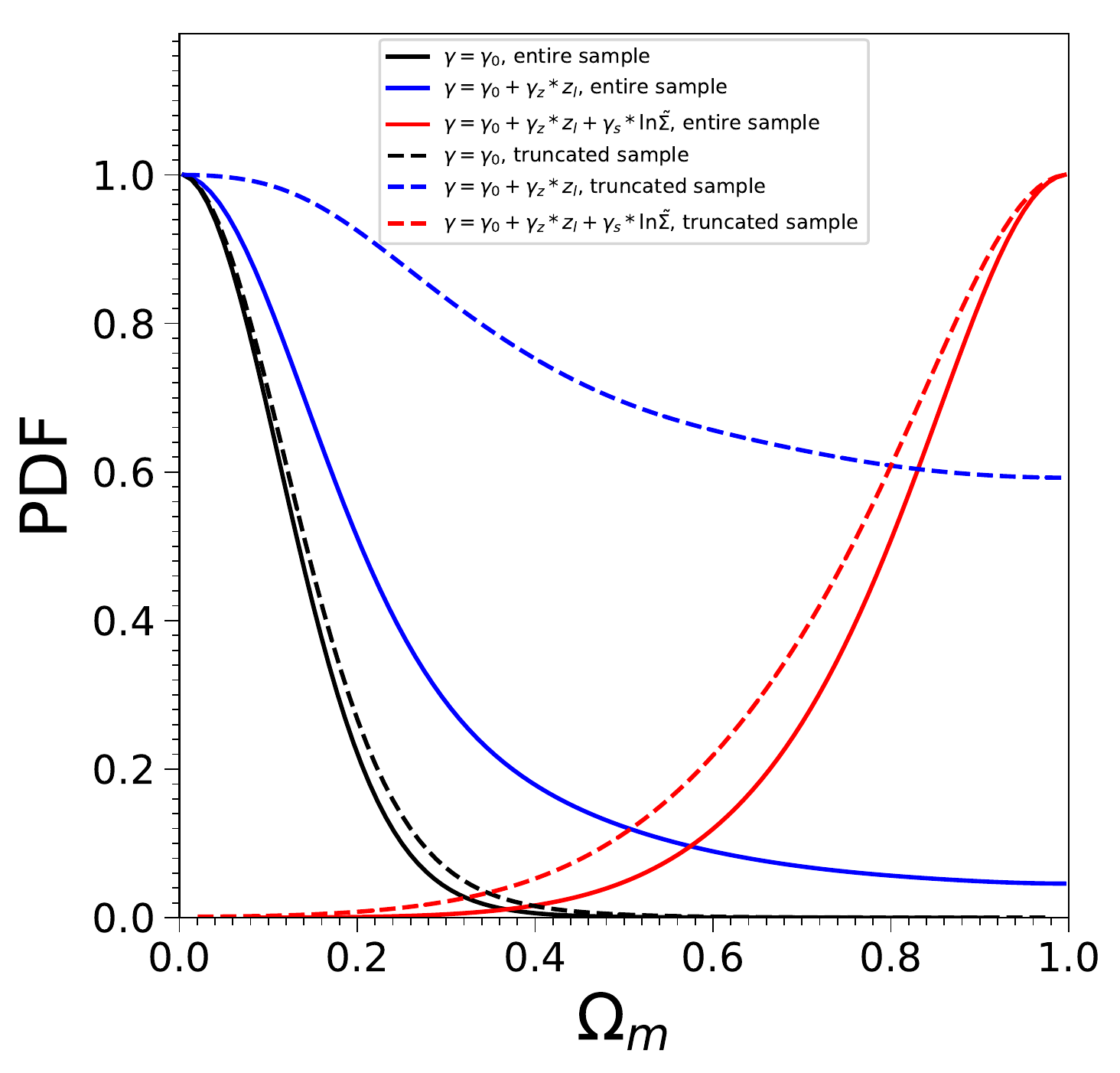}
 \caption{The 1D marginalized distributions of $\Omega_m$ constrained from the entire sample with 161 SGL systems (solid lines) and from the truncated sample with 130 systems (dashed lines), respectively, where $\delta$ (i.e., the
 logarithmic slope of the luminosity density profile) is treated as a universal parameter for all lens galaxies in the sample,
  and three different parameterizations for $\gamma$ (i.e., the logarithmic slope of the total density profile) are considered.}
\label{fig:om_1d}
\end{figure*}

The results corresponding to three parameterizations for $\gamma$ are displayed in Figure {\ref{fig:om_1d}}. The limits on $\Omega_m$ at 68\% (95\%) confidence level are $\Omega_m < 0.067$($\Omega_m < 0.158$), $\Omega_m < 0.227$($\Omega_m < 0.731$) and $\Omega_m>0.832$ ($\Omega_m>0.616$) in the scenarios of parameterizing $\gamma$ with ``$P_1$'', ``$P_2$'' and ``$P_3$'', respectively. The main tendencies can be summarized to three aspects.
First, the limits on $\Omega_m$ are significantly dependent on the lens mass model. The allowed range of $\Omega_m$ in the third scenario is inconsistent with those obtained in the former two at 68\% confidence level.
Second, the constraints on $\Omega_m$ are weak.  In the first two scenarios, the lower limits on $\Omega_m$ are unavailable. Conversely , the upper limit on $\Omega_m$ is absent in the last scenario. It is consistent with the qualitative analysis mentioned previously, which reveals that the sample under consideration is insensitivity to $\Omega_m$.  Third, the estimations on $\Omega_m$ are biased. The mean value of $\Omega_m$ constrained from the standard cosmological probes is around 0.3 (see, e.g. Huterer \& Shafer 2018;  Scolnic et al. 2018; Alam et al. 2017), such as $\Omega_m = 0.315\pm0.007$ in the framework of  flat $\Lambda$CDM model obtained from the recent Planck 2018 result (Planck Collaboration: Aghanim et al. 2018). In the first scenario, the limit on $\Omega_m$ is inconsistent with that from the Placnck result at 95\% confidence level, and the allowed values of $\Omega_m$ are especially low. In the second scenario, it is consistent with the Placnk result at 95\% confidence level, but the mean value of $\Omega_m$ is much lower. In the last scenario, it is inconsistent with the Placnk result at 95\% confidence level, and the mean value is much higher.

This is the first time to constrain the cosmological parameter in the scenario of considering the dependence of $\gamma$ on both redshift and surface mass density. In C15, they constrained the equation of state (EoS) of dark energy (with other cosmological parameters fixed) from their sample with 118 systems in the scenarios of $P_1$ and $P_2$. From the results listed in Table 2 of C15, one can see that the constraints on the EoS of dark energy are also quite weak, the uncertainties are bigger than 30\%.

\subsubsection{The case of treating $\delta$ as an observable for each lens galaxy}
\begin{figure*}
\includegraphics[width=0.95\linewidth]{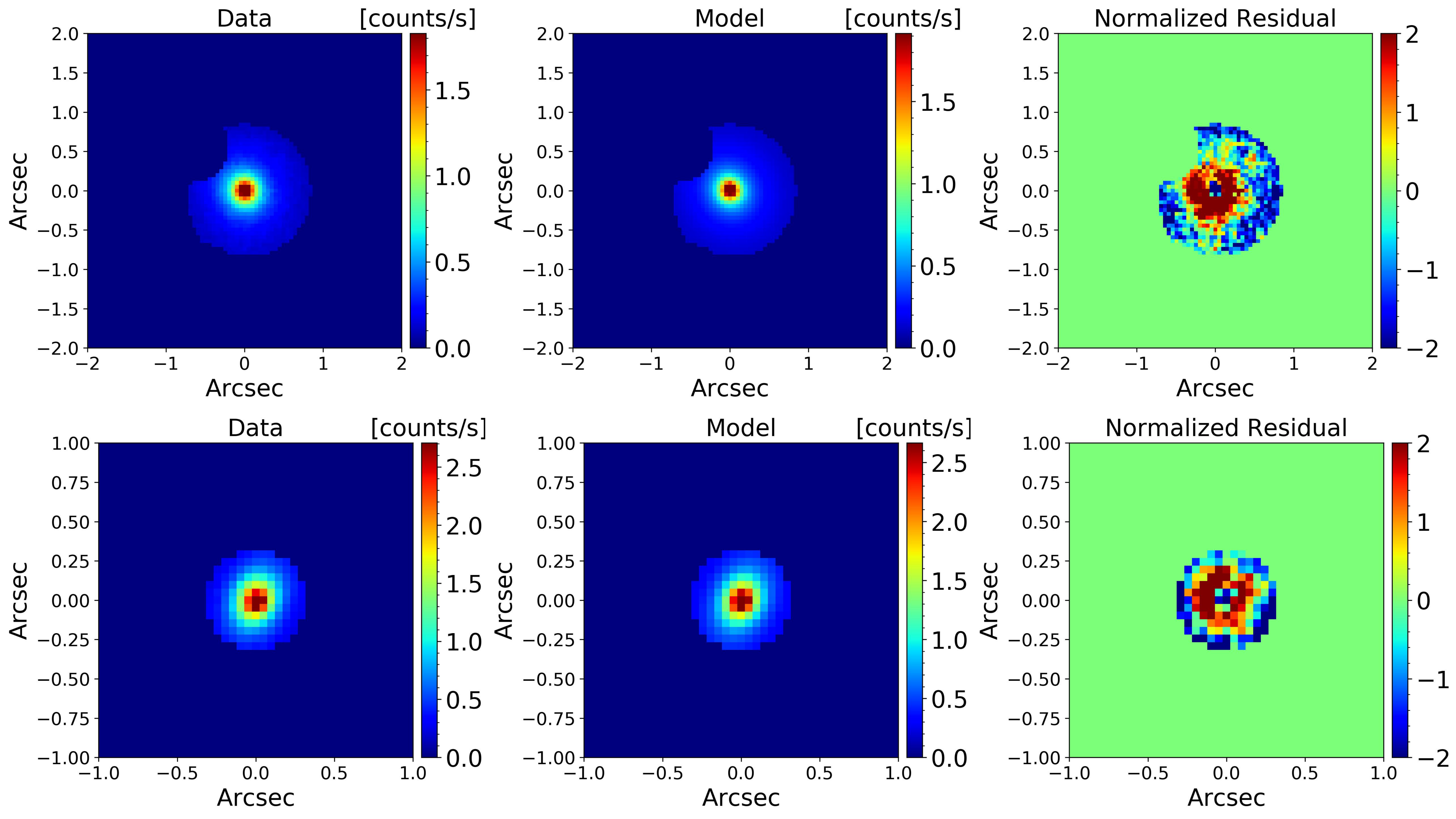}
\caption{Fitting the luminosity distributions over the aperture with radius $\theta_{\textrm{eff}}/2$, for lens galaxies,  where a power-law luminosity distribution model is assumed. SLACSJ0008$-$0004 (first row) and SDSSJ1215$+$0047 (second row) are taken as examples.
Each row shows the observed HST image (left panel), the reconstructed model for
surface brightness distribution of the lens galaxy (middle panel), and the normalized residual (right panel)
representing a difference between the reconstructed model and the observed image, which is dimensionless and defined as (Data-Model)/Noise.}
\label{fig:LuminosityFitting}
\end{figure*}

In the previous analyses, the most troublesome issue  is the bias in the estimation of $\Omega_m$, which may be due to some unconsidered systematic errors. The sample dependence of $\delta$ has been ignored in the above analysis, that may be a potential source of bias in the estimation of $\Omega_m$. To verify this conjecture, we choose to consider the intrinsic scatter of $\delta$ among the lens galaxies by treating $\delta$ as an observable for each lens galaxy.
We obtain $\delta$ values by fitting the two-dimensional power-law luminosity profile convolved with the instrumental point-spread function (PSF)
to HST F814W or F606W imaging data over a circle of radius $\theta_{\textrm{eff}}/2$  centered on the lens galaxies\footnote{The observational and model-predicted values of the velocity dispersion (i.e., Eqs. (\ref{eq:sigma_obs}) and (\ref{eq:sigma_th})) used in the analysis are the components of luminosity-weighted average over the aperture with radius $\theta_{\textrm{eff}}/2$, so here we do the luminosity fitting inside $\theta_{\textrm{eff}}/2$ for each lens galaxy.}, where the projected two-dimensional profile, $I(R)\propto R^{-\delta+1}$, is derived from the corresponding three-dimensional profile, i.e., $\nu(r)\propto r^{-\delta}$ (Bolton et al. 2006b and Schwab et al. 2010).
Obviously, the high-resolution HST imaging data must be available for the selected lens galaxies. In view of this requirement,
a truncated sample with 130 SGL systems is used here, where the chosen systems come from the SLACS, S4TM, BELLS and BELLS GALLERY surveys.
According to the actually available imaging data, we use the HST F814W data for the SLACS, S4TM, and BELLS lenses, and HST F606W data for BELLS GALLERY lenses.  It is worth pointing out that the region of interest in our method is the innermost area for each lens galaxy, wherein
the luminosity density profile can be well fitted with a power-law distribution. The Figure \ref{fig:LuminosityFitting} shows the luminosity fitting results, where SLACSJ0008$-$0004 from SLACS and SDSSJ1215$+$0047 from BELLS are
 taken as examples.  It turns out that the statistical error on the measured $\delta$
for each lens galaxy is smaller than 1\%, which is ignorable.
Besides, the measured values of $\delta$ for the 130 lenses have a mean of $<\delta> = 2.173$ and a standard deviation $\sigma_{\delta}=0.085$.

\begin{figure*}
\includegraphics[width=0.95\linewidth]{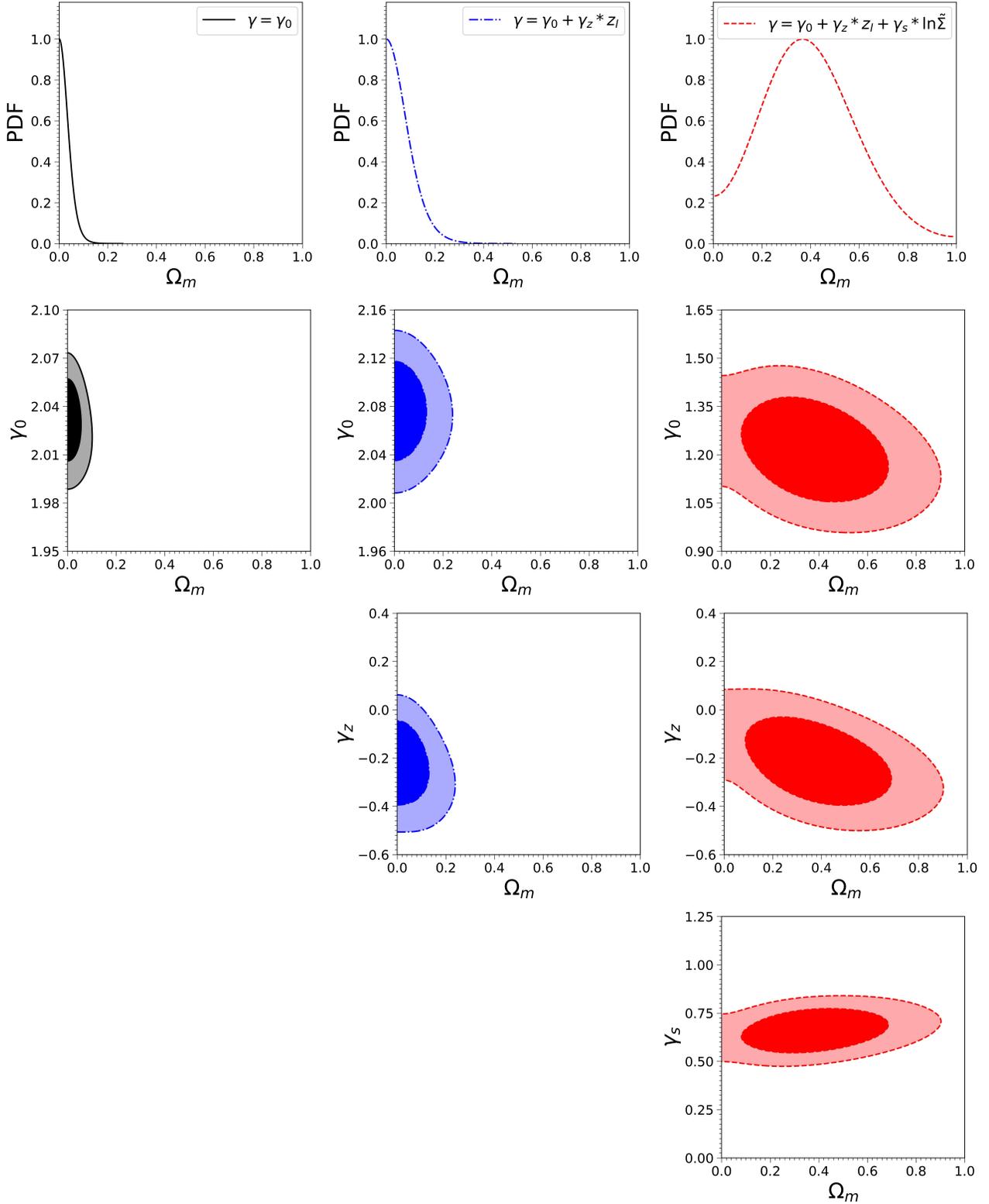}
 \caption{The 1D and 2D probability distributions of model parameters constrained from the truncated sample with 130 SGL systems,
 where the HST F814W or F606W imaging data for these lens galaxies are available from the archive. After obtaining the logarithmic slopes of the luminosity density profiles for each lens from the imaging data, we consider three different parameterizations for the logarithmic slope ($\gamma$) of the total density profile.
 The columns, from left to right, correspond to the scenarios  $\gamma = \gamma_0$, $\gamma = \gamma_0 + \gamma_z*z_l$ and $\gamma = \gamma_0 + \gamma_z*z_l+\gamma_s \log \tilde{\Sigma}$, respectively.}
\label{fig:combine_130_delta_ob}
\end{figure*}

The observational constraints on the free parameters are presented in Figure \ref{fig:combine_130_delta_ob} and Table \ref{tab:results}. It turns out that the estimation of  $\Omega_m$ is still obviously biased in the scenarios of parameterizing $\gamma$ with $P_1$ and $P_2$. On the contrary,
the unbiased estimate of $\Omega_m$ is obtained in the scenario of parameterizing $\gamma$ with $P_3$,
though the uncertainty on $\Omega_m$ is about 50\%.
To make a comparison, the constraints on $\Omega_m$ from the truncated sample in the case of treating $\delta$ as a universal parameter are also displayed in Figure \ref{fig:om_1d},  which show that the estimations on $\Omega_m$ are still biased like those from the entire sample. It implies that the unbiased estimate on $\Omega_m$
in the scenario of parameterizing $\gamma$ with $P_3$ results from treating $\delta$ as an observable for each lens rather than reducing the sample from 161 to 130 systems.
Hence, in order to get the unbiased estimate for $\Omega_m$, one should properly consider
 both the dependence of $\gamma$ on the lens properties and the intrinsic scatter of $\delta$ among the lenses.
Moreover, when replacing the previously adopted prior value on $\beta$ with $\beta = 0$, the estimation of $\Omega_m$ is shifted
 from $\Omega_m=0.381^{+0.185}_{-0.154}$ to $\Omega_m = 0.176^{+0.134}_{-0.101}$ at 68\% confidence level in the scenario of $P_3$, these two estimations are consistent at 68\% confidence level, but the relative change in the mean value is $\sim$50\%. The effect of the prior on $\beta$ is noticeable, so a precise prior on $\beta$ is very important.

The analyses above reveal that the limit on the cosmological parameter is quite dependent on the lens mass model. So it is necessary to compare the lens mass models and select the most compatible one, that can supply helpful reference for future studies on selecting the lens mass model.  First of all, it is easy to imagine that the lens model which can result in an unbiased estimate of $\Omega_m$ should be preferred. In order to ensure the rigorousness of the consequence, we employ the Bayesian information criterion (BIC) to compare the lens models. The BIC (Schwarz 1978) is defined as
\begin{equation}
\label{eq:BIC}
\textrm{BIC}=-2\ln \mathcal{L}_{max}+k\ln N,
\end{equation}
where $\mathcal{L}_{max}$ is the maximum likelihood (satisfying $-2\ln \mathcal{L}_{max}=\chi^2_{min}$ under the Gaussian assumption), $k$ is the number of the parameters of the considered model, and $N$ is the number of data points used in the fitting. The BIC is widely used in a cosmological context(see, e.g., Liddle 2004; God{\l}owski \& Szyd{\l}owski 2005;
Magueijo \& Sorkin 2007; Mukherjee et al. 2006; Biesiada 2007; Davis et al. 2007; Li et al. 2013; Wen et al. 2018; Birrer et al. 2019). This statistic prefers
models that give a good fit with fewer parameters. The favorite model is the one with the minimum BIC value.
The BIC values for the scenarios of parameterizing $\gamma$ with $P_1$, $P_2$ and $P_3$ are 334.7, 332.6 and 207.5, respectively, which are also listed in the last column of Table \ref{tab:results}.
So, the most compatible lens model is the third scenario,  which is exactly the one that results in an unbiased estimate of $\Omega_m$.

What's more, in the framework of the third scenario, $\gamma_z=0$ is ruled out at $\sim$$2\sigma$ level, and  $\gamma_s = 0$ is ruled out at $\sim$$10\sigma$ level, wherein $\gamma_z = -0.218^{+0.089}_{-0.087}$
and $\gamma_s = 0.661^{+0.054}_{-0.055}$ at 68\% confidence level.
By fixing the cosmological parameters at the fiducial values, Sonnenfeld et al. (2013a) found the dependence of $\gamma$ on redshift and surface stellar mass density ($\Sigma_{\ast}$) at  $\sim$$3.1\sigma$ and $\sim$$5.4\sigma$ levels  from the SL2S, SLACS and LSD lenses, wherein $\partial \gamma/\partial z_l=-0.31^{+0.09}_{-0.10}$ and
$\partial \gamma/\partial \log\Sigma_{\ast}=0.38\pm0.07$ at 68\% confidence level.
 Based on the previous analysis, we conclude that dependencies of $\gamma$ on both the redshift and the surface mass density are significant. Besides, $\gamma$ has a positive correlation with the surface mass density,
 and a negative correlation with the redshift.

 \section{Summary and conclusions}\label{sec:Conclusions}
We have compiled a galaxy-scale strong gravitational lensing sample including 161 systems with the gravitational lensing and stellar velocity dispersion measurements, which are selected with strict criteria to satisfy the assumption of spherical symmetry on the lens mass model. Actually, the selected lenses are all early-type galaxies with E/S0 morphologies. A kind of spherically symmetric mass distributions expressed with Eq.(\ref{eq:profile}) is assumed for the lens galaxies throughout this paper.
After carrying out the qualitative and semi-quantitative analysis, we find that the current sample is weak at confining cosmological parameters.
Besides, the slope of the total mass density profile, i.e., $\gamma$, presents a significant sample-dependent signal. On the other side,
the sample dependence of the slope of the luminosity density, i.e., $\delta$, is much weaker than that of $\gamma$, but stronger than that of the orbit anisotropy parameter $\beta$.
Given this, we specifically consider three parameterizations for the slope $\gamma$.
 The slope $\gamma$ is treated as an arbitrary constant without considering any dependency in the first scenario (namely ``P1''). And its dependence on the lens redshift is considered in the second scenario (namely ``P2''). Further, its dependencies on both the redshift and the surface mass density of the lens are taken into account in the last scenario  (namely ``P3'').  Moreover, $\beta$ is treated as a nuisance parameter and has been marginalized over with a Gaussian prior $\beta = 0.18\pm0.13$ from the independent constraint based on the observations of nearby elliptical galaxies.
 Regarding to the parameter $\delta$, we treat it in two different ways.
In the first case, we treat $\delta$ as a universal parameter for all lens galaxies in the entire sample.
  It turns out that the limit on the cosmological parameter, $\Omega_m$, is quite weak and biased, as well as quite dependent on the
  parametrization of $\gamma$.
In the second case, we turn to consider the sample-dependence of $\delta$ by treating $\delta$ as an observable for each lens.
Then, the observational constraints show that the unbiased estimate of $\Omega_m$ can be obtained in the scenario of parameterizing $\gamma$
with $P_3$, although the estimates are still biased in the scenarios of $P_1$ and $P_2$.
The dependencies of $\gamma$ on the redshift and on the surface mass density are observed at $\sim$$2\sigma$ and $\sim$$10\sigma$ levels, respectively.

Consequently, both the dependence of $\gamma$ on the lens properties and the intrinsic scatter of $\delta$ among
 the lenses should be properly taken into account to get the unbiased estimate for the cosmological parameter in the method under consideration. Besides, the effect of the prior on $\beta$ is also noticeable, so a empirically-motivated prior on $\beta$ is very essential for our study.
 It also shows that the slope $\gamma$ has a positive correlation with the surface mass density, and a negative correlation with the redshift. The overall trends show that, at a given redshift, the galaxies with high density also have steeper slopes; and, at fixed surface mass density, the galaxies at a lower redshift have steeper slopes. These trends are consistent with those obtained in the previous studies (e.g., Auger et al. 2010;  Ruff et al. 2011;  Bolton et al 2012;  Holanda et al. 2017; Sonnenfeld et al. 2013a; Li et al. 2018b).
 It is worth noting that what's measured here is how the mean density slope
for the population of ETGs considered changes in the $(z_l, \tilde{\Sigma})$ space, and not how $\gamma$ changes along the
lifetime of an individual galaxy. In order to infer
the latter quantity one needs to evaluate the variation of $\gamma$
along the evolutionary track of a galaxy as it moves
in the $(z_l, \tilde{\Sigma})$ space. This requires to know how
both mass and size of a galaxy change with time (Sonnenfeld et al. 2013a), since
the slope depends on these parameters, however, that is impractical in the actual observations. Hence, the numerical simulations are usually needed in order to obtain the evolutionary track of an individual galaxy.
Finally, we point out that besides the dependence of $\gamma$ on redshift and surface mass density considered in this work, other important dependencies may also be found in future, that can lead to a more accurate phenomenological model for lens galaxies.

In addition, although the measurements of the velocity dispersions ($\sigma$) of lens galaxies alone are weak at constraining the cosmological parameters, measurements of time delays ($\Delta \tau$) and the joint measurements of the former two ($\Delta \tau/\sigma^2$) are both proved to be more sensitive to the cosmological parameters (Paraficz \& Hjorth 2009; Wei \& Wu 2017; Jee et al. 2015, 2016; Shajib et al. 2018).  At present, the cosmological implementation is limited by the uncertainty in the lens modeling.
One can anticipate a dramatic increase in the number of SGL systems in view of the forthcoming optical imaging surveys (see, e.g., Oguri \& Marshall 2010; Collett 2015; Shu et al. 2018). The high-quality imaging and spectroscopic observations on SGL systems will certainly be very helpful in improving the lens mass modelling (Barnab\`e et al. 2013; Suyu et al. 2017; Treu et al. 2018). In addition, the mass and surface-brightness structures of early-type lensing galaxies may be better quantified with the aid of hydrodynamic simulations of galaxy formation n (Xu et al. 2017; Mukherjee et al. 2018; Wang et al. 2018).

\section*{Acknowledgments}
We would like to thank the anonymous referee for her/his comments that helped to greatly improve and clarify several points of the paper, and thank Adam S. Bolton for the helpful explanation on how to fit the luminosity distribution with a single power-law profile,
and thank Shuo Cao, Marek Biesiada, Kai Liao, Xuheng Ding, Lixin Xu, Guojian Wang and Jingzhao Qi for other helpful discussions.
YC has been supported by the National Natural Science Foundation of China (Nos. 11703034 and 11573031), and the NAOC Nebula Talents Program.
RL has been supported by the National Key Program for Science and Technology Research and Development of China (2017YFB0203300), the National Natural Science Foundation of China (Nos. 11773032 and 118513), and the NAOC Nebula Talents Program. YS has been supported by the Royal Society -- K.C. Wong International Fellowship (NF170995).

%%%%%%%%%%%%%%%%%%%%%%%%%%%%%%%%%%%%%%%%%%%%%%%%%%%%%%%%%%%%%%%%%%%%%%%%%%%%%%
%%%%%%%%%%%%%%%%%%%%%%%%%%%%%%%%%%%%%%%%%%%%%%%%%%%%%%%%%%%%%%%%%%%%%%%%%%%%%%

\clearpage
\begin{table*}
\begin{center}%{\scriptsize%\footnotesize
\caption{\label{tab:results}  The 1D marginalized limits for model parameters constrained from the truncated sample with 130 SGL systems,
where the employed observational data and lens mass models are the same as those involved in Figure \ref{fig:combine_130_delta_ob}.
All limits and confidence regions quoted here are 68\%.}
\begin{tabular}{lllllll}\hline\hline
Lens Model & Parameters &  &  & &$\chi^2_{\textrm{min}}/\textrm{d.o.f}$ & BIC   \\
\hline
$\gamma=\gamma_0$ & $\Omega_m<0.023$ & $\gamma_0 = 2.030^{+0.011}_{-0.012}$ & ... & ...& $325/130$ & 334.7 \\
$\gamma=\gamma_0+\gamma_z*z_l$ &$\Omega_m<0.064$ & $\gamma_0 = 2.076^{+0.018}_{-0.019}$ & $\gamma_z=-0.235^{+0.077}_{-0.083}$ & ...& $318/130$ & 332.6\\
$\gamma=\gamma_0+\gamma_z*z_l+\gamma_s*\log\tilde{\Sigma}$& $\Omega_m=0.381^{+0.185}_{-0.154}$& $\gamma_0 = 1.213^{+0.078}_{-0.079}$ & $\gamma_z=-0.218^{+0.089}_{-0.087}$ & $\gamma_s=0.661^{+0.054}_{-0.055}$& $188/130$ & 207.5\\
\hline
\end{tabular}
\end{center}
\end{table*}

\clearpage

\appendix
\onecolumn
\section{Observational data of the selected galaxy-scale SGL systems}~\label{sec:sample}

 %\begin{table*}
\begin{center}
\begin{longtable}{lllccrcclc}
%%\centering
%%\begin{threeparttable}
%%\caption{\label{tab:results}Effect of Trade Openness on Environment (Air Pollution)}
\caption[]{The selected galaxy-scale SGL systems}\label{tab:sample}\\
%%\begin{threeparttable}
%
%%\begin{supertabular}{lcclcccccl}%\hline\hline
%%\begin{longtable}{lcclcccccl}%\hline\hline
%%\toprule
\hline
\hline
\multicolumn{1}{c}{Lens Name} & \multicolumn{1}{c}{$z_l$} & \multicolumn{1}{c}{$z_s$}  & \multicolumn{1}{c}{$\theta_E[^{\prime\prime}]$} & \multicolumn{1}{c}{$\theta_{\textrm{eff}}[^{\prime\prime}]$} & \multicolumn{1}{c}{slit$ [^{\prime\prime}\times ^{\prime\prime}]$} &	 \multicolumn{1}{c}{Fiber radius $[^{\prime\prime}]$}  &	 \multicolumn{1}{c}{$\theta_{\textrm{ap}}[^{\prime\prime}]$}  &	 \multicolumn{1}{c}{$\sigma_{ap}[km/s]$} &	\multicolumn{1}{c}{Survey Name}\\
%%\midrule
\hline
\endfirsthead
\multicolumn{10}{c} {\tablename\ \thetable{} }\\
\multicolumn{10}{c}{(continued)}\\
%%\toprule
\hline
\multicolumn{1}{c}{Lens Name} & \multicolumn{1}{c}{$z_l$} & \multicolumn{1}{c}{$z_s$}  & \multicolumn{1}{c}{$\theta_E[^{\prime\prime}]$} & \multicolumn{1}{c}{$\theta_{\textrm{eff}}[^{\prime\prime}]$} & \multicolumn{1}{c}{slit$ [^{\prime\prime}\times ^{\prime\prime}]$} &	 \multicolumn{1}{c}{Fiber radius $[^{\prime\prime}]$}  &	 \multicolumn{1}{c}{$\theta_{\textrm{ap}}[^{\prime\prime}]$}  &	 \multicolumn{1}{c}{$\sigma_{ap}[km/s]$} &	\multicolumn{1}{c}{Survey Name}\\
%%\midrule
\hline
\endhead
%
%\bottomrule
%\multicolumn{10}{c}{{(-- continued on next page)}}
\hline
\hline
\endfoot
%\bottomrule
%%{NOTE--Columns are: 1) survey name; 2) the number of data points from the corresponding survey; 3-4)
%%the range of the lens redshifts (col. 3) and their median value (col. 4);
%% 5-6) the range of the central velocity dispersions (col. 5) and their median value (col. 6), in [km/s].}
\hline
\hline
\endlastfoot

MG2016$+$112	&	1.004 	&	3.263 	&	1.56 	&	0.31 	&	$1\times1.25$	&	...	&	0.65	&	 $304\pm27$	&	LSD	\\
0047$-$281	&	0.485 	&	3.595 	&	1.34 	&	0.82 	&	$0.4\times 1.25$	&	...	&	0.41	&	 $219\pm12$	&	LSD	\\
CFRS03.1077	&	0.938 	&	2.941 	&	1.24 	&	1.60 	&	$0.5\times1.25$	&	...	&	0.46	&	$256\pm19$	 &	LSD	\\
HST14176$+$5226	&	0.810 	&	3.399 	&	1.41 	&	1.06 	&	$0.32\times1.25$	&	...	&	0.37	&	 $212\pm18$	&	LSD	\\
HSTT15433$+$5352	&	0.497 	&	2.092 	&	0.36 	&	0.41 	&	$0.3\times1.25$	&	...		&	0.35	&	 $108\pm14$	&	LSD	\\
SL2SJ020524$-$93023	&	0.557 	&	1.330 	&	0.76 	&	0.75 	&	$0.9\times 1.60$	&	...		&	0.69 	 &	$276\pm37$	&	SL2S	 \\
SL2SJ021247$-$055552	&	0.750 	&	2.740 	&	1.27 	&	1.22 	&	$0.9\times 1.60$	&	...	&	0.69 	 &	$273\pm22$	&	SL2S	 \\
SL2SJ021325$-$074355	&	0.717 	&	3.480 	&	2.39 	&	1.97 	&	$1.0\times 1.68$	&	...	&	0.75 	 &	$293\pm34$	&	SL2S	 \\
SL2SJ021411$-$040502	&	0.609 	&	1.880 	&	1.41 	&	1.21 	&	$1.0\times 1.88$	&	...	&	0.79 	 &	$287\pm47$	&	SL2S	 \\
SL2SJ021737$-$051329	&	0.646 	&	1.850 	&	1.27 	&	0.73 	&	$1.5\times 1.68$	&	...	&	0.92 	 &	$239\pm27$	&	SL2S	 \\
SL2SJ021801$-$080247	&	0.884 	&	2.060 	&	1.00 	&	1.02 	&	$0.9\times 1.60$	&	...	&	0.69 	 &	$246\pm48$	&	SL2S	 \\
SL2SJ021902$-$082934	&	0.389 	&	2.150 	&	1.30 	&	0.95 	&	$1.0 \times 1.68$	&	...	&	0.75 	 &	$289\pm23$	&	SL2S	 \\
SL2SJ022046$-$094927	&	0.572 	&	2.610 	&	1.00 	&	0.53 	&	$1.0 \times 1.90$	&	...	&	0.80 	 &	$254\pm29$	&	SL2S	 \\
SL2SJ022511$-$045433	&	0.238 	&	1.200 	&	1.76 	&	2.12 	&	$1.0 \times 0.81$	&	...	&	0.52 	 &	$234\pm21$	&	SL2S	 \\
SL2SJ022610$-$042011	&	0.494 	&	1.230 	&	1.19 	&	0.84 	&	$1.0 \times 1.62$	&	...	&	0.74 	 &	$263\pm24$	&	SL2S	 \\
SL2SJ023251$-$040823	&	0.352 	&	2.340 	&	1.04 	&	1.14 	&	$1.0\times 1.68$	&	...	&	0.75 	 &	$281\pm26$	&	SL2S	 \\
SL2SJ084847$-$035103	&	0.682 	&	1.550 	&	0.85 	&	0.45 	&	$0.9\times 1.60$	&	...	&	0.69 	 &	$197\pm21$	&	SL2S	 \\
SL2SJ084909$-$041226	&	0.722 	&	1.540 	&	1.10 	&	0.46 	&	$0.9\times 1.60$	&	...	&	0.69 	 &	$320\pm24$	&	SL2S	 \\
SL2SJ084959$-$025142	&	0.274 	&	2.090 	&	1.16 	&	1.34 	&	$0.9\times 1.60$	&	...	&	0.69 	 &	$276\pm35$	&	SL2S	 \\
SL2SJ085540$-$014730	&	0.365 	&	3.390 	&	1.03 	&	0.69 	&	$0.7 \times 1.62$	&	...	&	0.62 	 &	$222\pm25$	&	SL2S	 \\
SL2SJ090407$-$005952	&	0.611 	&	2.360 	&	1.40 	&	2.00 	&	$0.9\times 1.60$	&	...	&	0.69 	 &	$183\pm21$	&	SL2S	 \\
SL2SJ095921$+$020638	&	0.552 	&	3.350 	&	0.74 	&	0.46 	&	$0.9\times 1.60$	&	...	&	0.69 	 &	$188\pm22$	&	SL2S	 \\
SL2SJ135949$+$553550	&	0.783 	&	2.770 	&	1.14 	&	1.13 	&	$1.0\times 1.62$	&	...	&	0.74 	 &	$228\pm29$	&	SL2S	 \\
SL2SJ140454$+$520024	&	0.456 	&	1.590 	&	2.55 	&	2.03 	&	$1.0\times 1.62$	&	...	&	0.74 	 &	$342\pm20$	&	SL2S	 \\
SL2SJ140546$+$524311	&	0.526 	&	3.010 	&	1.51 	&	0.83 	&	$1.0\times 1.62$	&	...	&	0.74 	 &	$284\pm21$	&	SL2S	 \\
SL2SJ140650$+$522619	&	0.716 	&	1.470 	&	0.94 	&	0.80 	&	$1.0\times 1.62$	&	...	&	0.74 	 &	$253\pm19$	&	SL2S	 \\
SL2SJ141137$+$565119	&	0.322 	&	1.420 	&	0.93 	&	0.85 	&	$1.0\times 1.62$	&	...	&	0.74 	 &	$214\pm23$	&	SL2S	 \\
SL2SJ142059$+$563007	&	0.483 	&	3.120 	&	1.40 	&	1.62 	&	$1.0\times 1.62$	&	...	&	0.74 	 &	$228\pm19$	&	SL2S	 \\
SL2SJ220329$+$020518	&	0.400 	&	2.150 	&	1.95 	&	0.99 	&	$1.0\times 1.62$	&	...	&	0.74 	 &	$213\pm21$	&	SL2S	 \\
SL2SJ220506$+$014703	&	0.476 	&	2.530 	&	1.66 	&	0.66 	&	$0.9\times 1.60$	&	...	&	0.69 	 &	$317\pm30$	&	SL2S	 \\
SL2SJ222148$+$011542	&	0.325 	&	2.350 	&	1.40 	&	1.12 	&	$1.0\times 1.88$	&	...	&	0.79 	 &	$222\pm23$	&	SL2S	 \\

SDSSJ0008$-$0004	&	0.440 	&	1.192 	&	1.16 	&	1.71 	&	...	&	1.5	&	1.5	&	$193\pm36$	&	 SLACS	\\
SDSSJ0029$-$0055	&	0.227 	&	0.931 	&	0.96 	&	2.16 	&	...	&	1.5	&	1.5	&	$229\pm18$	&	 SLACS	\\
SDSSJ0037$-$0942	&	0.195 	&	0.632 	&	1.53 	&	2.19 	&	...	&	1.5	&	1.5	&	$279\pm10$	&	 SLACS	\\
SDSSJ0044$+$0113	&	0.120 	&	0.197 	&	0.79 	&	2.61 	&	...	&	1.5	&	1.5	&	$266\pm13$	&	 SLACS	\\
SDSSJ0109$+$1500	&	0.294 	&	0.525 	&	0.69 	&	1.38 	&	...	&	1.5	&	1.5	&	$251\pm19$	&	 SLACS	\\
SDSSJ0157$-$0056	&	0.513 	&	0.924 	&	0.79 	&	1.06 	&	...	&	1.5	&	1.5	&	$295\pm47$	&	 SLACS	\\
SDSSJ0216$-$0813	&	0.332 	&	0.523 	&	1.16 	&	2.67 	&	...	&	1.5	&	1.5	&	$333\pm23$	&	 SLACS	\\
SDSSJ0252$+$0039	&	0.280 	&	0.982 	&	1.04 	&	1.39 	&	...	&	1.5	&	1.5	&	$164\pm12$	&	 SLACS	\\
SDSSJ0330$-$0020	&	0.351 	&	1.071 	&	1.10 	&	1.20 	&	...	&	1.5	&	1.5	&	$212\pm21$	&	 SLACS	\\
SDSSJ0405$-$0455	&	0.075 	&	0.810 	&	0.80 	&	1.36 	&	...	&	1.5	&	1.5	&	$160\pm7$	&	 SLACS	\\
SDSSJ0728$+$3835	&	0.206 	&	0.688 	&	1.25 	&	1.78 	&	...	&	1.5	&	1.5	&	$214\pm11$	&	 SLACS	\\
SDSSJ0737$+$3216	&	0.322 	&	0.581 	&	1.00 	&	2.82 	&	...	&	1.5	&	1.5	&	$338\pm16$	&	 SLACS	\\
SDSSJ0822$+$2652	&	0.241 	&	0.594 	&	1.17 	&	1.82 	&	...	&	1.5	&	1.5	&	$259\pm15$	&	 SLACS	\\
SDSSJ0903$+$4116	&	0.430 	&	1.065 	&	1.29 	&	1.78 	&	...	&	1.5	&	1.5	&	$223\pm27$	&	 SLACS	\\
SDSSJ0912$+$0029	&	0.164 	&	0.324 	&	1.63 	&	3.87 	&	...	&	1.5	&	1.5	&	$326\pm12$	&	 SLACS	\\
SDSSJ0935$-$0003	&	0.347 	&	0.467 	&	0.87 	&	4.24 	&	...	&	1.5	&	1.5	&	$396\pm35$	&	 SLACS	\\
SDSSJ0936$+$0913	&	0.190 	&	0.588 	&	1.09 	&	2.11 	&	...	&	1.5	&	1.5	&	$243\pm11$	&	 SLACS	\\
SDSSJ0946$+$1006	&	0.222 	&	0.609 	&	1.38 	&	2.35 	&	...	&	1.5	&	1.5	&	$263\pm21$	&	 SLACS	\\
SDSSJ0956$+$5100	&	0.241 	&	0.470 	&	1.33 	&	2.19 	&	...	&	1.5	&	1.5	&	$334\pm15$	&	 SLACS	\\
SDSSJ0959$+$4416	&	0.237 	&	0.531 	&	0.96 	&	1.98 	&	...	&	1.5	&	1.5	&	$244\pm19$	&	 SLACS	\\
SDSSJ0959$+$0410	&	0.126 	&	0.535 	&	0.99 	&	1.39 	&	...	&	1.5	&	1.5	&	$197\pm13$	&	 SLACS	\\
SDSSJ1016$+$3859	&	0.168 	&	0.439 	&	1.09 	&	1.46 	&	...	&	1.5	&	1.5	&	$247\pm13$	&	 SLACS	\\
SDSSJ1020$+$1122	&	0.282 	&	0.553 	&	1.20 	&	1.59 	&	...	&	1.5	&	1.5	&	$282\pm18$	&	 SLACS	\\
SDSSJ1023$+$4230	&	0.191 	&	0.696 	&	1.41 	&	1.77 	&	...	&	1.5	&	1.5	&	$242\pm15$	&	 SLACS	\\
SDSSJ1029$+$0420	&	0.104 	&	0.615 	&	1.01 	&	1.56 	&	...	&	1.5	&	1.5	&	$210\pm9$	&	 SLACS	\\
SDSSJ1100$+$5329	&	0.317 	&	0.858 	&	1.52 	&	2.24 	&	...	&	1.5	&	1.5	&	$187\pm23$	&	 SLACS	\\
SDSSJ1106$+$5228	&	0.095 	&	0.407 	&	1.23 	&	1.68 	&	...	&	1.5	&	1.5	&	$262\pm9$	&	 SLACS	\\
SDSSJ1112$+$0826	&	0.273 	&	0.629 	&	1.49 	&	1.50 	&	...	&	1.5	&	1.5	&	$320\pm20$	&	 SLACS	\\
SDSSJ1134$+$6027	&	0.153 	&	0.474 	&	1.10 	&	2.02 	&	...	&	1.5	&	1.5	&	$239\pm11$	&	 SLACS	\\
SDSSJ1142$+$1001	&	0.222 	&	0.504 	&	0.98 	&	1.91 	&	...	&	1.5	&	1.5	&	$221\pm22$	&	 SLACS	\\
SDSSJ1143$-$0144	&	0.106 	&	0.402 	&	1.68 	&	4.80 	&	...	&	1.5	&	1.5	&	$269\pm5$	&	 SLACS	\\
SDSSJ1153$+$4612	&	0.180 	&	0.875 	&	1.05 	&	1.16 	&	...	&	1.5	&	1.5	&	$226\pm15$	&	 SLACS	\\
SDSSJ1204$+$0358	&	0.164 	&	0.631 	&	1.31 	&	1.47 	&	...	&	1.5	&	1.5	&	$267\pm17$	&	 SLACS	\\
SDSSJ1205$+$4910	&	0.215 	&	0.481 	&	1.22 	&	2.59 	&	...	&	1.5	&	1.5	&	$281\pm13$	&	 SLACS	\\
SDSSJ1213$+$6708	&	0.123 	&	0.640 	&	1.42 	&	3.23 	&	...	&	1.5	&	1.5	&	$292\pm11$	&	 SLACS	\\
SDSSJ1218$+$0830	&	0.135 	&	0.717 	&	1.45 	&	3.18 	&	...	&	1.5	&	1.5	&	$219\pm10$	&	 SLACS	\\
SDSSJ1250$+$0523	&	0.232 	&	0.795 	&	1.13 	&	1.81 	&	...	&	1.5	&	1.5	&	$252\pm14$	&	 SLACS	\\
SDSSJ1402$+$6321	&	0.205 	&	0.481 	&	1.35 	&	2.70 	&	...	&	1.5	&	1.5	&	$267\pm17$	&	 SLACS	\\
SDSSJ1403$+$0006	&	0.189 	&	0.473 	&	0.83 	&	1.46 	&	...	&	1.5	&	1.5	&	$213\pm17$	&	 SLACS	\\
SDSSJ1416$+$5136	&	0.299 	&	0.811 	&	1.37 	&	1.43 	&	...	&	1.5	&	1.5	&	$240\pm25$	&	 SLACS	\\
SDSSJ1420$+$6019	&	0.063 	&	0.535 	&	1.04 	&	2.06 	&	...	&	1.5	&	1.5	&	$205\pm4$	&	 SLACS	\\
SDSSJ1430$+$4105	&	0.285 	&	0.575 	&	1.52 	&	2.55 	&	...	&	1.5	&	1.5	&	$322\pm32$	&	 SLACS	\\
SDSSJ1436$-$0000	&	0.285 	&	0.805 	&	1.12 	&	2.24 	&	...	&	1.5	&	1.5	&	$224\pm17$	&	 SLACS	\\
SDSSJ1443$+$0304	&	0.134 	&	0.419 	&	0.81 	&	0.94 	&	...	&	1.5	&	1.5	&	$209\pm11$	&	 SLACS	\\
SDSSJ1451$-$0239	&	0.125 	&	0.520 	&	1.04 	&	2.48 	&	...	&	1.5	&	1.5	&	$223\pm14$	&	 SLACS	\\
SDSSJ1525$+$3327	&	0.358 	&	0.717 	&	1.31 	&	2.90 	&	...	&	1.5	&	1.5	&	$264\pm26$	&	 SLACS	\\
SDSSJ1531$-$0105	&	0.160 	&	0.744 	&	1.71 	&	2.50 	&	...	&	1.5	&	1.5	&	$279\pm12$	&	 SLACS	\\
SDSSJ1538$+$5817	&	0.143 	&	0.531 	&	1.00 	&	1.58 	&	...	&	1.5	&	1.5	&	$189\pm12$	&	 SLACS	\\
SDSSJ1621$+$3931	&	0.245 	&	0.602 	&	1.29 	&	2.14 	&	...	&	1.5	&	1.5	&	$236\pm20$	&	 SLACS	\\
SDSSJ1627$-$0053	&	0.208 	&	0.524 	&	1.23 	&	1.98 	&	...	&	1.5	&	1.5	&	$290\pm14$	&	 SLACS	\\
SDSSJ1630$+$4520	&	0.248 	&	0.793 	&	1.78 	&	1.96 	&	...	&	1.5	&	1.5	&	$276\pm16$	&	 SLACS	\\
SDSSJ1636$+$4707	&	0.228 	&	0.675 	&	1.09 	&	1.68 	&	...	&	1.5	&	1.5	&	$231\pm15$	&	 SLACS	\\
SDSSJ2238$-$0754	&	0.137 	&	0.713 	&	1.27 	&	2.33 	&	...	&	1.5	&	1.5	&	$198\pm11$	&	 SLACS	\\
SDSSJ2300$+$0022	&	0.228 	&	0.463 	&	1.24 	&	1.83	&	...	&	1.5	&	1.5	&	$279\pm17$	&	 SLACS	\\
SDSSJ2303$+$1422	&	0.155 	&	0.517 	&	1.62 	&	3.28	&	...	&	1.5	&	1.5	&	$255\pm16$	&	 SLACS	\\
SDSSJ2321$-$0939	&	0.082 	&	0.532 	&	1.60 	&	4.11	&	...	&	1.5	&	1.5	&	$249\pm8$	&	 SLACS	\\
SDSSJ2341$+$0000	&	0.186 	&	0.807 	&	1.44 	&	3.15	&	...	&	1.5	&	1.5	&	$207\pm13$	&	 SLACS	\\
SDSSJ0143$-$1006	&	0.2210 	&	1.1046 	&	1.23 	&	3.24 	&	...	&	1.5	&	1.5	&	$203\pm17$	&	 S4TM	\\
SDSSJ0159$-$0006	&	0.1584 	&	0.7477 	&	0.92 	&	1.58 	&	...	&	1.5	&	1.5	&	$216\pm18$	&	 S4TM	\\
SDSSJ0324$+$0045	&	0.3210 	&	0.9199 	&	0.55 	&	1.67 	&	...	&	1.5	&	1.5	&	$183\pm19$	&	 S4TM	\\
SDSSJ0324$-$0110	&	0.4456 	&	0.6239 	&	0.63 	&	2.23 	&	...	&	1.5	&	1.5	&	$310\pm38$	&	 S4TM	\\
SDSSJ0753$+$3416	&	0.1371 	&	0.9628 	&	1.23 	&	1.89 	&	...	&	1.5	&	1.5	&	$208\pm12$	&	 S4TM	\\
SDSSJ0754$+$1927	&	0.1534 	&	0.7401 	&	1.04 	&	1.46 	&	...	&	1.5	&	1.5	&	$193\pm16$	&	 S4TM	\\
SDSSJ0757$+$1956	&	0.1206 	&	0.8326 	&	1.62 	&	3.67 	&	...	&	1.5	&	1.5	&	$206\pm11$	&	 S4TM	\\
SDSSJ0826$+$5630	&	0.1318 	&	1.2907 	&	1.01 	&	1.64 	&	...	&	1.5	&	1.5	&	$163\pm8$	&	 S4TM	\\
SDSSJ0847$+$2348	&	0.1551 	&	0.5327 	&	0.96 	&	1.54 	&	...	&	1.5	&	1.5	&	$199\pm16$	&	 S4TM	\\
SDSSJ0851$+$0505	&	0.1276 	&	0.6371 	&	0.91 	&	1.35 	&	...	&	1.5	&	1.5	&	$175\pm11$	&	 S4TM	\\
SDSSJ0920$+$3028	&	0.2881 	&	0.3918 	&	0.70 	&	4.25 	&	...	&	1.5	&	1.5	&	$297\pm17$	&	 S4TM	\\
SDSSJ0955$+$3014	&	0.3214 	&	0.4671 	&	0.54 	&	2.95 	&	...	&	1.5	&	1.5	&	$271\pm33$	&	 S4TM	\\
SDSSJ0956$+$5539	&	0.1959 	&	0.8483 	&	1.17 	&	1.96 	&	...	&	1.5	&	1.5	&	$188\pm11$	&	 S4TM	\\
SDSSJ1010$+$3124	&	0.1668 	&	0.4245 	&	1.14 	&	3.26 	&	...	&	1.5	&	1.5	&	$221\pm11$	&	 S4TM	\\
SDSSJ1041$+$0112	&	0.1006 	&	0.2172 	&	0.60 	&	2.50 	&	...	&	1.5	&	1.5	&	$200\pm7$	&	 S4TM	\\
SDSSJ1048$+$1313	&	0.1330 	&	0.6679 	&	1.18 	&	1.90 	&	...	&	1.5	&	1.5	&	$195\pm10$	&	 S4TM	\\
SDSSJ1051$+$4439	&	0.1634 	&	0.5380 	&	0.99 	&	1.66 	&	...	&	1.5	&	1.5	&	$216\pm16$	&	 S4TM	\\
SDSSJ1056$+$4141	&	0.1343 	&	0.8318 	&	0.72 	&	1.81 	&	...	&	1.5	&	1.5	&	$157\pm10$	&	 S4TM	\\
SDSSJ1101$+$1523	&	0.1780 	&	0.5169 	&	1.18 	&	0.89 	&	...	&	1.5	&	1.5	&	$270\pm15$	&	 S4TM	\\
SDSSJ1116$+$0729	&	0.1697 	&	0.6860 	&	0.82 	&	2.44 	&	...	&	1.5	&	1.5	&	$190\pm11$	&	 S4TM	\\
SDSSJ1127$+$2312	&	0.1303 	&	0.3610 	&	1.25 	&	2.69 	&	...	&	1.5	&	1.5	&	$230\pm9$	&	 S4TM	\\
SDSSJ1137$+$1818	&	0.1241 	&	0.4627 	&	1.29 	&	1.79 	&	...	&	1.5	&	1.5	&	$222\pm8$	&	 S4TM	\\
SDSSJ1142$+$2509	&	0.1640 	&	0.6595 	&	0.79 	&	1.51 	&	...	&	1.5	&	1.5	&	$159\pm10$	&	 S4TM	\\
SDSSJ1144$+$0436	&	0.1036 	&	0.2551 	&	0.76 	&	1.22 	&	...	&	1.5	&	1.5	&	$207\pm14$	&	 S4TM	\\
SDSSJ1213$+$2930	&	0.0906 	&	0.5954 	&	1.35 	&	1.73 	&	...	&	1.5	&	1.5	&	$232\pm7$	&	 S4TM	\\
SDSSJ1301$+$0834	&	0.0902 	&	0.5331 	&	1.00 	&	1.25 	&	...	&	1.5	&	1.5	&	$178\pm8$	&	 S4TM	\\
SDSSJ1330$+$1750	&	0.2074 	&	0.3717 	&	1.01 	&	2.85 	&	...	&	1.5	&	1.5	&	$250\pm12$	&	 S4TM	\\
SDSSJ1403$+$3309	&	0.0625 	&	0.7720 	&	1.02 	&	2.00 	&	...	&	1.5	&	1.5	&	$190\pm6$	&	 S4TM	\\
SDSSJ1430$+$6104	&	0.1688 	&	0.6537 	&	1.00 	&	2.24 	&	...	&	1.5	&	1.5	&	$180\pm15$	&	 S4TM	\\
SDSSJ1433$+$2835	&	0.0912 	&	0.4115 	&	1.53 	&	3.23 	&	...	&	1.5	&	1.5	&	$230\pm6$	&	 S4TM	\\
SDSSJ1541$+$3642	&	0.1406 	&	0.7389 	&	1.17 	&	1.55 	&	...	&	1.5	&	1.5	&	$194\pm11$	&	 S4TM	\\
SDSSJ1543$+$2202	&	0.2681 	&	0.3966 	&	0.78 	&	2.32 	&	...	&	1.5	&	1.5	&	$285\pm16$	&	 S4TM	\\
SDSSJ1550$+$2020	&	0.1351 	&	0.3501 	&	1.01 	&	1.68 	&	...	&	1.5	&	1.5	&	$243\pm9$	&	 S4TM	\\
SDSSJ1553$+$3004	&	0.1604 	&	0.5663 	&	0.84 	&	2.15 	&	...	&	1.5	&	1.5	&	$194\pm15$	&	 S4TM	\\
SDSSJ1607$+$2147	&	0.2089 	&	0.4865 	&	0.57 	&	2.63 	&	...	&	1.5	&	1.5	&	$197\pm16$	&	 S4TM	\\
SDSSJ1633$+$1441	&	0.1281 	&	0.5804 	&	1.39 	&	2.39 	&	...	&	1.5	&	1.5	&	$231\pm9$	&	 S4TM	\\
SDSSJ2309$-$0039	&	0.2905 	&	1.0048 	&	1.14 	&	2.08 	&	...	&	1.5	&	1.5	&	$184\pm13$	&	 S4TM	\\
SDSSJ2324$+$0105	&	0.1899 	&	0.2775 	&	0.59 	&	1.10 	&	...	&	1.5	&	1.5	&	$245\pm15$	&	 S4TM	\\
SDSSJ0801$+$4727	&	0.483 	&	1.518 	&	0.49 	&	0.50 	&	...	&	1	&	1	&	$98\pm24$	&	 BELLS	\\
SDSSJ1234$-$0241	&	0.490 	&	1.016 	&	0.53 	&	1.05 	&	...	&	1	&	1	&	$122\pm31$	&	 BELLS	\\
SDSSJ1352$+$3216	&	0.463 	&	1.034 	&	1.82 	&	0.58 	&	...	&	1	&	1	&	$161\pm21$	&	 BELLS	\\
SDSSJ1159$-$0007	&	0.579 	&	1.346 	&	0.68 	&	0.96 	&	...	&	1	&	1	&	$165\pm41$	&	 BELLS	\\
SDSSJ1318$-$0104	&	0.659 	&	1.396 	&	0.68 	&	0.69 	&	...	&	1	&	1	&	$177\pm27$	&	 BELLS	\\
SDSSJ1349$+$3612	&	0.440 	&	0.893 	&	0.75 	&	1.89 	&	...	&	1	&	1	&	$178\pm18$	&	 BELLS	\\
SDSSJ1221$+$3806	&	0.535 	&	1.284 	&	0.70 	&	0.47 	&	...	&	1	&	1	&	$187\pm48$	&	 BELLS	\\
SDSSJ0944$-$0147	&	0.539 	&	1.179 	&	0.73 	&	0.48 	&	...	&	1	&	1	&	$204\pm34$	&	 BELLS	\\
SDSSJ1601$+$2138	&	0.544 	&	1.446 	&	0.86 	&	0.44 	&	...	&	1	&	1	&	$207\pm36$	&	 BELLS	\\
SDSSJ1542$+$1629	&	0.352 	&	1.023 	&	1.04 	&	0.73 	&	...	&	1	&	1	&	$210\pm16$	&	 BELLS	\\
SDSSJ0151$+$0049	&	0.517 	&	1.364 	&	0.68 	&	0.67 	&	...	&	1	&	1	&	$219\pm39$	&	 BELLS	\\
SDSSJ1337$+$3620	&	0.564 	&	1.182 	&	1.39 	&	2.03 	&	...	&	1	&	1	&	$225\pm35$	&	 BELLS	\\
SDSSJ2125$+$0411	&	0.363 	&	0.978 	&	1.20 	&	0.90 	&	...	&	1	&	1	&	$247\pm17$	&	 BELLS	\\
SDSSJ1545$+$2748	&	0.522 	&	1.289 	&	1.21 	&	2.59 	&	...	&	1	&	1	&	$250\pm37$	&	 BELLS	\\
SDSSJ1215$+$0047	&	0.642 	&	1.297 	&	1.37 	&	0.65 	&	...	&	1	&	1	&	$262\pm45$	&	 BELLS	\\
SDSSJ0830$+$5116	&	0.530 	&	1.332 	&	1.14 	&	0.97 	&	...	&	1	&	1	&	$268\pm36$	&	 BELLS	\\
SDSSJ1631$+$1854	&	0.408 	&	1.086 	&	1.63 	&	1.43 	&	...	&	1	&	1	&	$272\pm14$	&	 BELLS	\\
SDSSJ2303$+$0037	&	0.458 	&	0.936 	&	1.02 	&	1.35 	&	...	&	1	&	1	&	$274\pm31$	&	 BELLS	\\
SDSSJ0747$+$4448	&	0.437 	&	0.897 	&	0.61 	&	0.92 	&	...	&	1	&	1	&	$281\pm52$	&	 BELLS	\\
SDSSJ2122$+$0409	&	0.626 	&	1.452 	&	1.58 	&	0.90 	&	...	&	1	&	1	&	$324\pm56$	&	 BELLS	\\
SDSSJ0747$+$5055	&	0.438 	&	0.898 	&	0.75 	&	1.09 	&	...	&	1	&	1	&	$328\pm60$	&	 BELLS	\\
SDSSJ0029$+$2544	&	0.5869 	&	2.4504 	&	1.34 	&	0.49 	&	...	&	1 	&	1	&	$241\pm45$	&	 BELLS GALLERY	\\
SDSSJ0201$+$3228	&	0.3957 	&	2.8209 	&	1.70 	&	2.32 	&	...	&	1 	&	1	&	$256\pm20$	&	 BELLS GALLERY	\\
SDSSJ0237$-$0641	&	0.4859 	&	2.2491 	&	0.65 	&	1.05 	&	...	&	1 	&	1	&	$290\pm89$	&	 BELLS GALLERY	\\
SDSSJ0742$+$3341	&	0.4936 	&	2.3633 	&	1.22 	&	0.89 	&	...	&	1 	&	1	&	$218\pm28$	&	 BELLS GALLERY	\\
SDSSJ0755$+$3445	&	0.7224 	&	2.6347 	&	2.05 	&	2.89 	&	...	&	1 	&	1	&	$272\pm52$	&	 BELLS GALLERY	\\
SDSSJ0856$+$2010	&	0.5074 	&	2.2335 	&	0.98 	&	0.51 	&	...	&	1 	&	1	&	$334\pm54$	&	 BELLS GALLERY	\\
SDSSJ0918$+$5104	&	0.5811 	&	2.4030 	&	1.60 	&	0.57 	&	...	&	1 	&	1	&	$298\pm49$	&	 BELLS GALLERY	\\
SDSSJ1110$+$2808	&	0.6073 	&	2.3999 	&	0.98 	&	1.45 	&	...	&	1 	&	1	&	$191\pm39$	&	 BELLS GALLERY	\\
SDSSJ1116$+$0915	&	0.5501 	&	2.4536 	&	1.03 	&	0.98 	&	...	&	1 	&	1	&	$274\pm55$	&	 BELLS GALLERY	\\
SDSSJ1141$+$2216	&	0.5858 	&	2.7624 	&	1.27 	&	0.44 	&	...	&	1 	&	1	&	$285\pm44$	&	 BELLS GALLERY	\\
SDSSJ1201$+$4743	&	0.5628 	&	2.1258 	&	1.18 	&	0.48 	&	...	&	1 	&	1	&	$239\pm43$	&	 BELLS GALLERY	\\
SDSSJ1226$+$5457	&	0.4980 	&	2.7322 	&	1.37 	&	0.56 	&	...	&	1 	&	1	&	$248\pm26$	&	 BELLS GALLERY	\\
SDSSJ2228$+$1205	&	0.5305 	&	2.8324 	&	1.28 	&	0.53 	&	...	&	1 	&	1	&	$255\pm50$	&	 BELLS GALLERY	\\
SDSSJ2342$-$0120	&	0.5270 	&	2.2649 	&	1.11 	&	1.75 	&	...	&	1 	&	1	&	$274\pm43$	&	 BELLS GALLERY	\\
\end{longtable}
{NOTE -- The columns are: (1) Lens name;(2) lens redshift;(3) source redshift;(4) Einstein angle;(5) effective radius;(6) angular sizes of width and
length of the rectangular aperture; (7) angular radius of the fiber;(8) angular radius of the circular aperture,
for the rectangular aperture the corresponding equivalent value derived with Eq. (\ref{eq:theta_ap_eff});
(9) velocity dispersion measured inside the circular aperture with the angular radius $\theta_{ap}$;(10) survey name.}
%%\end{threeparttable}
%\end{longtable}
\end{center}
%\end{table*}

\label{lastpage}

\end{document}